\documentclass[lettersize,journal]{IEEEtran}
\IEEEoverridecommandlockouts
\usepackage{url}
\usepackage{ifthen}
\usepackage{cite}
\usepackage{amsmath} 
\usepackage{amssymb}
\usepackage{graphicx}
\usepackage{epstopdf}
\usepackage{epsfig}
\usepackage{multirow}
\usepackage{amsthm}
\usepackage{comment}
\usepackage{caption}
\usepackage{enumitem}
\usepackage{color}
\usepackage{bbm}
\usepackage{algorithm}
\usepackage{algpseudocode}
\usepackage{csquotes}
\usepackage{subfiles}
\usepackage{caption, multirow, makecell}
\usepackage{array}
\usepackage{caption}
\usepackage{afterpage}
\usepackage{dblfloatfix}
\usepackage{hyperref}
\usepackage{graphicx}
\usepackage{balance}
\usepackage{subcaption}
\usepackage{diagbox}
\usepackage{float}

\newtheorem{thm}{Theorem}

\newtheorem{cor}{Corollary}
\newtheorem{example}{Example}

\newtheorem{defn}{Definition}

\newtheorem{rem}{Remark}
\def\BibTeX{{\rm B\kern-.05em{\sc i\kern-.025em b}\kern-.08em
		T\kern-.1667em\lower.7ex\hbox{E}\kern-.125emX}}

\begin{document}	
\title{Optimal Device-to-Device Placement Delivery Arrays using Combinatorial Designs
}
\author{\IEEEauthorblockN{Rashid Ummer N.T. and B. Sundar Rajan}\\
	\IEEEauthorblockA{Department of Electrical Communication Engineering, Indian Institute of Science, Bangalore, India \\
		E-mail: \{rashidummer, bsrajan\}@iisc.ac.in}
}
\maketitle	
	\begin{abstract}
Device-to-device (D2D) communication is one of the most promising techniques for fifth-generation and beyond wireless communication systems. This paper considers coded caching in a wireless D2D network, in which a central server initially places the data in the user cache memories, and all user demands are served through inter-user coded multicast transmissions. D2D placement delivery array (DPDA) was proposed as a tool for designing coded caching schemes with reduced subpacketization levels in a D2D network. In this paper, we first constructed three classes of DPDAs using a cross resolvable design, a group divisible design, and a newly developed block design. The resulting D2D schemes achieve low subpacketization levels while meeting the known lower bound on the transmission load of a DPDA. These classes of constructed DPDAs either simplify or generalize all existing DPDA constructions that achieve the known lower bound and have low subpacketization levels. Furthermore, a new lower bound on the transmission load of a DPDA is proposed. Two new classes of DPDAs are then constructed using a cross resolvable design and a newly developed block design, respectively. These constructions yield low-subpacketization D2D schemes and achieve the proposed lower bound on the transmission load. Compared to existing schemes with the same system parameters as those obtained from the proposed DPDAs, the proposed schemes have an advantage in either transmission load or subpacketization level or both.
	\end{abstract}

	\section{Introduction}
		Wireless data traffic is increasing rapidly every year, driven predominantly by video-on-demand \cite{Eri}. {\footnotetext{A part of the content has been presented at the IEEE WCNC 2025 held in Milan, Italy, March 2025 (doi: 10.1109/WCNC61545.2025.10978210).}} Cache-aided communication exploits the high temporal variability in network traffic. The novel coded caching scheme proposed by Maddah-Ali and Niesen in \cite{MaN} (referred to as the MAN scheme) considered an error-free broadcast network consisting of a single server with a library of $N$ files connected to a set of $K$ users, each having a cache of size $M$ files. The content is placed in the caches during off-peak times, such that the demands of all users are met using multicast coded transmissions and cache contents. The MAN scheme is optimal under uncoded data placement when $N \geq K$ \cite{YMA}. To achieve the optimal transmission load of the MAN scheme, each file needs to be split into $F$ packets, referred to as the subpacketization level, which grows exponentially with the number of users for a given memory ratio $\frac{M}{N}$.
		
		Device-to-device (D2D) communication is one of the most promising techniques for next-generation wireless Internet of Things (IoT) networks \cite{MACF, ERW}. D2D communication is also an enabling technology for realizing IoT in future fifth-generation and beyond wireless cellular communication systems \cite{TUY, LLLW}.  Low latency is an essential constraint for many emerging IoT applications, and edge caching in IoT networks is a potential solution for decreasing the latency \cite{ZTS}. Therefore, studying cache-aided D2D networks is of great relevance to IoT networks. Ji, Caire, and Molisch in \cite{Ji} extended the coded caching scheme to a wireless D2D network consisting of $K$ users each having a cache of size $M$ files and a central server with a library of $N$ files (referred to as the JCM scheme). During the placement phase, the server places uncoded data into the user caches from the library. During the delivery phase, no central server is present and each user broadcasts coded messages using their cache content to all other users through an error-free shared medium. By using multicast coded transmissions and cached content, all users receive their requested files. The order optimality of the transmission load of the JCM scheme was shown in \cite{Ji} and \cite{CKRG}. Similar to the MAN scheme, the JCM scheme requires a subpacketization level that grows exponentially with the number of users for a given memory ratio $\frac{M}{N}$. For practical scenarios, it is desirable to have D2D coded caching schemes with low subpacketization levels.
		
		 In \cite{YCT}, Yan \textit{et al.} proposed an array called a Placement Delivery Array (PDA), which describes both the placement and delivery phases of a centralized coded caching system.  Motivated by this, Wang \textit{et al.} in \cite{JMQX} proposed an array called the D2D placement delivery array (DPDA), which characterizes the placement and delivery phases in a D2D coded caching system. Low subpacketization level D2D coded caching schemes can be obtained by constructing the appropriate DPDAs. D2D coded caching schemes obtained from a DPDA are with uncoded data placement and one-shot delivery. A lower bound on the transmission load of the D2D scheme for a given $K,\frac{M}{N} \text{ and } F$ from a DPDA was derived in \cite{JMQX}, and it was shown that the JCM scheme achieves the lower bound. The three classes of DPDAs given in \cite{JMQX_Arxiv} also achieve the lower bound on the transmission load, while requiring a lower subpacketization level than that of the JCM scheme. Two of these three classes of DPDAs were constructed recursively. No other known D2D scheme achieves the lower bound for the transmission load. Obtaining new lower bounds and developing direct constructions of low subpacketization level D2D schemes that achieve lower bounds on the transmission load are explored in this paper. 
		 
	The contributions of this paper are as follows: \\
	\noindent $\bullet$ Three new classes of DPDAs that achieve the known lower bound on the transmission load of a DPDA are constructed using combinatorial designs. The D2D coded caching schemes obtained from these classes of DPDAs achieve the same transmission load as the JCM scheme, but with lower subpacketization levels. Among these DPDAs, the first class of DPDA (Theorem \ref{thm:dpda1}) is obtained from a class of maximal cross resolvable design (MCRD) specified in Corollary \ref{cor:CRD}. One DPDA construction in \cite{JMQX_Arxiv} has the same parameters as the DPDA in Theorem \ref{thm:dpda1}; however, it was not obtained using CRDs. Moreover, the DPDA construction in Theorem \ref{thm:dpda1} is generalized to other classes of CRDs to obtain new classes of DPDAs (Corollary \ref{cor:conc1_gen}). The second class of DPDA (Theorem \ref{thm:dpda4}) is obtained from the class of group divisible design specified in Corollary \ref{triv_GDD}. The third class of DPDA (Theorem \ref{thm:dpda5}) is obtained from the combinatorial design described in Section \ref{sec_constrn3}. In  \cite{JMQX_Arxiv}, two classes of DPDAs with the same parameters as those of Theorem \ref{thm:dpda4} and Theorem \ref{thm:dpda5} were constructed recursively, starting from smaller arrays. The proposed constructions are direct constructions that use combinatorial designs, thereby greatly simplifying existing recursive constructions.    
		
	\noindent $\bullet$ A new lower bound on the transmission load of a DPDA is obtained (Theorem \ref{thm:lb_dpda}). 
	
	\noindent $\bullet$  Two novel classes of DPDAs that achieve the proposed lower bound on the transmission load of a DPDA are constructed using combinatorial designs. Among these DPDAs, the first class of DPDA (Theorem \ref{thm:dpda2}) is obtained from a class of MCRD specified in Corollary \ref{cor:CRD}, and the second class of DPDA (Theorem \ref{thm:dpda3}) is obtained using the combinatorial design described in Section \ref{constrn_dpda5}. These classes of DPDAs provide D2D schemes with a subpacketization level that grows sub-linearly with the number of users.
	
	\noindent $\bullet$ Compared to existing schemes with the same system parameters as those obtained from the proposed DPDAs, the proposed schemes have an advantage in either transmission load or subpacketization level or both.
	
	The rest of this paper is organized as follows. In Section \ref{prelim}, we review the system model, preliminaries of DPDA, known D2D schemes, and useful definitions from combinatorial designs. The existing and proposed lower bounds on the transmission load of a DPDA are discussed in Section \ref{lb_dpda}. Novel DPDA constructions using specific classes of combinatorial designs are discussed in Section \ref{constr_d2d}, where each construction is followed by its performance analysis by comparing it with existing D2D coded caching schemes and by showing the optimality of each resulting DPDA. Finally, Section \ref{concl_d2d} concludes the paper.
	
	\textit{Notations}: For any positive integer $n$, $[n]$ denotes set $\{1,2,...,n\}$. For any integers $i$ and $n$ such that $0 \le i \le n$, $\binom{n}{i}$ denotes the binomial coefficient, which is calculated as $\frac{n!}{i!(n-i)!}$, $i \mid n$ reads as $i$ divides $n$, and $i \nmid n$ reads as $i$ does not divide $n$. For any set $\mathcal{A}$, $|\mathcal{A}|$ denotes the cardinality of $\mathcal{A}$. For a set $\mathcal{A}$ and a positive integer $i \leq |\mathcal{A}|$,  $\binom{\mathcal{A}}{i}$ denotes all the $i$-sized subsets of $\mathcal{A}$. For sets $\mathcal{A} \text{ and }\mathcal{B}$, $\mathcal{A}\textbackslash \mathcal{B}$ denotes the elements in $\mathcal{A}$ but not in $\mathcal{B}$. 
	
	\section{Preliminaries}\label{prelim}
		In this section, we first describe the D2D coded caching network model and then review the concepts of PDA and DPDA. Known D2D coded caching schemes are summarized next. We then review some basic definitions and properties of the combinatorial designs used in this paper.	
	\subsection{D2D Coded caching network model}
\begin{figure}[!htbp]
	\centering
	\captionsetup{justification=centering}
	\includegraphics[width=0.475\textwidth]{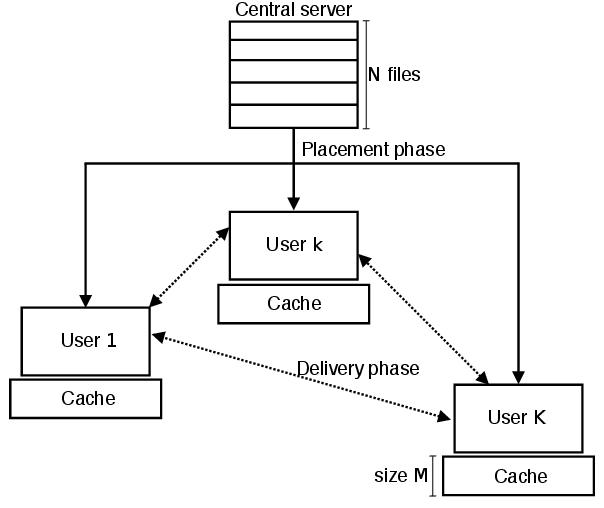}
	\caption{D2D coded caching network model.}
	\label{fig:settingd2d}
	\end{figure}

	A $(K, M, N)$ D2D coded caching network, depicted in Fig.\ref{fig:settingd2d}, consists of a server with a library of $N$ files $\{W_n: n \in [N]\}$ each of size $B$ bits and $K$ users each having a cache of size $M$ files where $M<N$. We assume $M\geq \frac{N}{K}$ to ensure that any possible demands can be met using the user cache contents. 
	
	The coded caching scheme operates in two phases. In the \textit{placement phase}, each file $W_{n}$ is split into $F$ non-overlapping packets and each user $k \in [K]$ places $MB$ bits in its cache, denoted by $\mathcal{Z}_k$. The data placement is assumed to be uncoded, in which case $MB$ bits are taken directly from the total $NB$ bits in the library. During the \textit{delivery phase}, each user requests a file from $\{W_n: n \in [N]\}$. The set of files requested by the users is represented by a demand vector $\vec{d}=(d_1,\ldots,d_K)$. For a given demand vector $\vec{d}$, each user broadcasts a coded message $X_{k,\vec{d}}$, of size $S_{k,\vec{d}}$ packets, using its cache content to all other users through the error-free shared medium. We assume the delivery to be a \textit{one-shot delivery}, that is, the user can recover any bit of any requested file using its cache content and at most one transmission. Using these transmissions and individual cache contents, each user obtains the requested file. The corresponding transmission load $R$ normalized to the file size is given by
	$ R \triangleq \max_{\vec{d}} \frac{\sum_{k=1}^{K}S_{k,\vec{d}}}{F}$.	
	 
	Next we review the concepts of PDA and DPDA.
	\subsection{PDA and DPDA}
	\begin{defn}
		(\hspace{1sp}\cite{YCT}) For positive integers $K, F, Z$ and $S$, an $F \times K$ array $\mathbf{P}=(p_{j,k})$, $j \in [F]$ and $k \in [K]$, composed of a specific symbol $\star$ and $S$ non-negative integers $[S]$, is called a $(K, F, Z, S)$ placement delivery array (PDA) if it satisfies the following conditions: \\
		\textit{C1}. The symbol $\star$ appears $Z$ times in each column.\\
		\textit{C2}. Each integer occurs at least once in the array.\\
		\textit{C3}. For any two distinct entries $p_{j_1,k_1}$ and $p_{j_2,k_2}$,\\ $p_{j_1,k_1}=p_{j_2,k_2}=s$ is an integer only if (\textit{a}) $j_1 \neq j_2$, $k_1 \neq k_2$, i.e., they lie in distinct rows and distinct columns, and (\textit{b}) $p_{j_1,k_2}=p_{j_2,k_1}=\star$.

	\end{defn} 
	If each integer appears $g$ times in $\mathbf{P}$ where $g$ is a constant, then $\mathbf{P}$ is said to be a $g$-regular $(K, F, Z, S)$ PDA or $g-(K, F, Z, S)$ PDA.
		\begin{defn}\label{def_dpda}
		(\hspace{1sp}\cite{JMQX}) A $(K, F, Z, S)$ PDA $\mathbf{P}=(p_{j,k})$, $j \in [F]$ and $k \in [K]$ is called a $(K, F, Z, S)$ DPDA if the additional property described below is satisfied. \\
		\textit{C4}. There is a mapping $\phi$ from $[S]$ to the users $[K]$ such that if $p_{j,k}=s$ for any $s \in [S]$, then $p_{j,\phi(s)}=\star$.
	\end{defn} 
	Given a DPDA, a D2D coded caching scheme can be obtained as explained in Theorem \ref{thm:d2d}. 
	\begin{thm}\label{thm:d2d}
		(\hspace{1sp}\cite{JMQX}) For a given $(K, F, Z, S)$ DPDA $\mathbf{P}=(p_{j,k})_{F \times K}$, a $(K,M,N)$ D2D coded caching scheme can be obtained with subpacketization $F$ and $\frac{M}{N}=\frac{Z}{F}$ using Algorithm \ref{d2d_alg}. For any demand $\vec{d}$, the demands of all users are satisfied with a transmission load of $R=\frac{S}{F}$.
	\end{thm}
		\begin{algorithm}[h]
		\renewcommand{\thealgorithm}{1}
		\caption{D2D coded caching scheme based on DPDA~\protect\cite{JMQX}}
		\label{d2d_alg}
		\begin{algorithmic}[1]
			\Procedure{Placement}{$\mathbf{P},\mathcal{W}$}       
			\State Split each file $W_n$ in $\mathcal{W}$ into $F$ packets: $W_n =\{W_{n,j}: j \in [F]\}$
			\For{\texttt{$k \in [K]$}}
			\State  $\mathcal{Z}_k$ $\leftarrow$ $\{W_{n,j}: p_{j,k}=\star, \forall n \in [N]\}$
			\EndFor
			\EndProcedure
			\Procedure{Delivery}{$\mathbf{P},\mathcal{W},\phi,\vec{d}$} 
			\For{\texttt{$s \in [S]$}}
			\State User $\phi(s)$ sends $\underset{p_{j,k}=s, j\in [F],k\in[K]}{\bigoplus}W_{d_k,j}$
			\EndFor    
			\EndProcedure
		\end{algorithmic}
	\end{algorithm}
	Example \ref{eg:dpda} illustrates Algorithm \ref{d2d_alg} and Theorem \ref{thm:d2d}.
	\begin{example}\label{eg:dpda}
It is easy to see that the following array $\mathbf{P}$ is a $2-(6,4,2,6)$ DPDA. Mapping $\phi(s)$ is the identity function.  
\begin{equation*}	
	\mathbf{P}	=	\begin{array}{cc|*{6}{c}}
		 & & 1 & 2 & 3	& 4 & 5 & 6  \\
		\hline
		&	1 & \star & 3 &\star & 5 & \star & 1 \\	
		&	2 & \star & 6 & 1 & \star & 4 & \star \\
		&	3 & 3 & \star & \star & 6 & 2 & \star \\
		&	4 & 5 & \star & 2 & \star & \star & 4 \\
	\end{array}
\end{equation*}
		Based on this DPDA $\mathbf{P}$, one can obtain a $(6,3,6)$ coded caching scheme with $F=4$ using Algorithm \ref{d2d_alg}, as follows: \\
		\noindent $\bullet$ \textbf{Placement phase:} From Line $2$ of Algorithm \ref{d2d_alg}, each file $W_n, \forall n \in [6]$ is split into $4$ packets, i.e., $W_n =\{W_{n,1},W_{n,2},W_{n,3},W_{n,4}\}$.  By lines $3-5$ of Algorithm \ref{d2d_alg}, each user caches the packets as follows:
			\begin{equation*}
	\begin{split}
		& \mathcal{Z}_{1} = \{W_{n,1},W_{n,2} \}, \mathcal{Z}_{2} = \{W_{n,3},W_{n,4} \},  \mathcal{Z}_{3} = \{W_{n,1},W_{n,3} \}, \\ & \mathcal{Z}_{4} = \{W_{n,2},W_{n,4} \},  \mathcal{Z}_{5} = \{W_{n,1},W_{n,4}\}, \mathcal{Z}_{6} = \{W_{n,2},W_{n,3}\} ,\\ & \forall n \in [6]. \\ 
	\end{split}
\end{equation*}
			\noindent $\bullet$ \textbf{Delivery phase:} Let the demand vector $\vec{d}=\{4,2,1,5,6,3\}$. By Lines $8-10$ of Algorithm \ref{d2d_alg}, users $1$ to $6$ transmit $W_{1,2} \oplus W_{3,1}, W_{1,4} \oplus W_{6,3}, W_{4,3} \oplus W_{2,1}, W_{6,2} \oplus W_{3,4}, W_{5,1} \oplus W_{4,4}$ and $W_{2,2} \oplus W_{5,3}$ respectively. Thus, each user can recover the requested file. For example, consider user $1$, whose demanded file is $W_4$. It has in its cache the packets $W_{4,1}$ and $W_{4,2}$ of the requested file $W_4$. From the transmission $W_{4,3} \oplus W_{2,1}$ by user $3$, user $1$ can recover $W_{4,3}$ since it has $W_{2,1}$ in its cache. From the transmission $W_{5,1} \oplus W_{4,4}$ by user $5$, user $1$ can recover $W_{4,4}$ because it has $W_{5,1}$ in its cache.  Thus, user $1$ gets all four packets of the requested file $W_4$.
		
		There are six transmissions in this example, one by each user, with each transmission being the size of a packet. Therefore, $R=\frac{6}{4}=1.5$ files, as indicated in Theorem \ref{thm:d2d}.
	\end{example} 
\begin{table*}[!htbp]
	\Large
	\setlength{\tabcolsep}{2.5pt}
	\resizebox{\linewidth}{!}{%
	\begin{tabular}{| c | c | c | c | c| c|c|}
		\hline
		\rule{0pt}{4ex}
		\makecell{Sl. \\ No.} &\makecell{Schemes} & Parameters & \makecell{ $K$} & \makecell{ $\frac{M}{N}$} & \makecell{$F$} & \makecell{ $R$}  \\ [4pt]	
		\hline
		\rule{0pt}{3.5ex}
		1 & \makecell{The JCM scheme in \cite{Ji}} & $K \in \mathbb{Z}^{+}, t \in [K-1]$  & $K$ & $\frac{t}{K}$ & $t\binom{K}{t}$ &  $\frac{K}{t}-1$ \\	
		\hline
		\rule{0pt}{3.5ex}
		2 & \makecell{Theorem 3 in \cite{JMQX} } &  $g+1-(K,F,Z,S)$ PDA & $K$ & $\frac{Z}{F}$  & $gF$ & $\frac{(g+1)S}{gF}$ \\
		\hline
		\rule{0pt}{3.5ex}
		3 & \makecell{Theorem 5 in \cite{JMQX_Arxiv} } &  any integer $q \ge 2$ & $2q$ & $\frac{1}{q}$  & $q^2$ & $q-1$ \\	
		\hline
		\rule{0pt}{3.5ex}
		4 & \makecell{Theorem 6 in \cite{JMQX_Arxiv} \\ (Recursive construction) } &  any integer $q \ge 2$ & $2q$ & $1-\frac{1}{q}$  & $2q(q-1)$ & $\frac{1}{q-1}$ \\	
		\hline
		\rule{0pt}{3.5ex}
		5 & \makecell{Theorem 7 in \cite{JMQX_Arxiv} \\ (Recursive construction)} &  any integer $q \ge 1$ & $2q+1$ & $1-\frac{2}{2q+1}$  & $4q^2-1$ & $\frac{2}{2q-1}$ \\	
		\hline
		\rule{0pt}{3.5ex}
		6 & \makecell{RS graph scheme in \cite{WCJ_RS} } & \makecell{Given $(\gamma,\tau)$-RS graph,\\ $\Lambda \in \mathbb{Z}^{+}, z \geq 2\Lambda$ } & $\Lambda^z$ & $2\Lambda^{\frac{-z}{2\Lambda^4 \ln \Lambda}}$  & $\Lambda^z(2\gamma-1)$ & $\frac{\tau}{\Lambda^z}\frac{2\gamma}{2\gamma-1}$ \\	
		\hline
		\rule{0pt}{3.5ex}
		7 & \makecell{Hypercube scheme in \cite{WCJ_HY} } &  $n \in \mathbb{Z}^{+}$ and $n \geq 2 $   & $n^2$ & $\frac{1}{n}$  & $n^n$ & $n$ \\
		\hline
		\rule{0pt}{3.5ex}
		8 & \makecell{Theorem 2 in \cite{JY} } &  \makecell{$n,a,t \in \mathbb{Z}^{+},$ \\ $ t \leq a \leq n-t $ }  & $\binom{n}{a}$ & $1-\frac{\binom{n-a}{t}\binom{a}{t}}{\binom{n}{2t}\binom{2t}{t}}$  & $\binom{n}{2t}\binom{2t}{t}$ & $\frac{\binom{n}{a}}{\binom{n}{2t}\binom{2t}{t}}$ \\
		\hline
		\rule{0pt}{3.5ex}
		9 & \makecell{Theorem 3 in \cite{JY} } &  $n,a \in \mathbb{Z}^{+}, n \geq 4a$  & $2\binom{n}{a}$ & $1-\frac{\binom{n-a}{a}}{\binom{n}{2a}}$  & $\binom{n}{2a}$ & $\frac{2\binom{n}{a}}{\binom{n}{2a}}$ \\	
		\hline
		\rule{0pt}{3.5ex}
		10 & \makecell{Theorem 4 in \cite{JY} } & \makecell{ $n$ is an odd integer, \\ $D$ is a $d$-subset of $\mathbb{Z}_{n}$ s.t. \\ $(i) \forall x \in D, \frac{x}{2} \notin D$;\\ $(ii) \forall$ distinct $x,y \in D, \frac{x+y}{2} \notin D$ }  & $n$ & $1-\frac{d}{n}$  & $n$ & $1$ \\	
		\hline
		\rule{0pt}{3.5ex}
		11 & \makecell{Theorem 5 in \cite{JY} } &  \makecell{$n,a,b \in \mathbb{Z}^{+},$ \\ $ a < b <2a < n $  } & $\binom{n}{a}$ & $1-\frac{\binom{a}{b-a}\binom{n-a}{a}}{\binom{n}{b}}$  & $\binom{n}{b}$ & $\frac{\binom{n}{a}\binom{n-a}{2a-b}}{\binom{n}{b}}$ \\
		\hline
		12 & Corollary 3 in \cite{MADCC_arxiv} & \makecell{$t-(v,k,\lambda)$ design,\\ $i \in [t-1]$ } & $\frac{\lambda \binom{v}{t}}{\binom{k}{t}}$ & $1-\frac{\binom{v-k}{i}}{\binom{v}{i}}$ & $\binom{v}{i}\binom{k}{t-i}$ & $\frac{\lambda\binom{v-i}{k}}{\binom{v-t}{k-t}\binom{t}{i}}$  \\ 
		\hline
		13 & Corollary 5 in \cite{MADCC_arxiv} & \makecell{$t-(m,q,k,1)$ GDD,\\ $s-(q,m,1)$ Orthogonal Array \\ with $1 \le t \le k \le s < m$, \\ $l \in [\min({m-s},{t-1})]$ } & $\frac{ \binom{m}{t}q^t}{\binom{k}{t}}$ & $1-\left(\frac{q-1}{q}\right)^k$  & $q^s\binom{k}{l}$ & $\le \frac{(q^m-q^s) q^{t-l} \binom{m-l}{t-l}}{q^s\binom{k}{l}\binom{k-l}{t-l}}$ \\ \hline
		14 & Theorem 12 in \cite{MADCC_arxiv} & \makecell{any positive integers $t$ and $m$ \\ such that $1<t<m$, and $q \geq 2$ } & $q^{m-1}$ & $1-\left(\frac{q-1}{q}\right)^t$  & $\binom{m}{t}q^t$ & $\frac{(q-1)^tq^{m-t-1}}{\binom{m}{t}}$ \\
		\hline
	\end{tabular}}	
	\caption{Known D2D coded caching schemes}	
	\label{tab:d2d_schemes}
\end{table*} 
	\subsection{Known D2D coded caching schemes}
	Table \ref{tab:d2d_schemes} summarizes the known D2D coded caching schemes. The first proposed scheme for a $(K, M, N)$ D2D network was the JCM scheme specified in Row $1$ of Table \ref{tab:d2d_schemes}. Given a $(g+1)-(K,F,Z,S)$ PDA, Wang \textit{et al.} in \cite{JMQX} proposed a method to construct a $(K,gF,gZ,(g+1)S)$ DPDA, as mentioned in Row $2$ of Table \ref{tab:d2d_schemes}. Since there are several PDA constructions discussed in the literature, one can obtain several DPDAs, but this construction will lead to schemes with higher subpacketization levels. The authors in \cite{JMQX} showed that by using the PDA corresponding to the MAN scheme, one can obtain the DPDA corresponding to the JCM scheme. Three classes of DPDAs in \cite{JMQX_Arxiv}, two of which were constructed recursively, achieves the same transmission load as the JCM scheme, while requiring a lower subpacketization level, are given in Rows $3$ to $5$ of Table \ref{tab:d2d_schemes}. 
	
	In \cite{WCJ_RS}, by using  Ruzsa-Szemeredi (RS) graphs, Woolsey \textit{et al.} proposed a D2D coded caching scheme with a subquadratic subpacketization level. The subpacketization level of the D2D coded caching scheme using the hypercube approach discussed in \cite{WCJ_HY} is exponentially less compared to that of the JCM scheme. The transmission loads of both these schemes were higher than those of the JCM scheme. Rows $6$ and $7$ of Table \ref{tab:d2d_schemes} summarize these two schemes. In \cite{JY}, Li and Chang provided four direct constructions of DPDAs which are summarized in Rows $8$ to $11$ of Table \ref{tab:d2d_schemes}. Several classes of D2D schemes were obtained from DPDAs constructed using combinatorial $t$-designs and $t$-group divisible designs in \cite{MADCC_arxiv}, which are summarized in Rows $12$ to $16$ of Table \ref{tab:d2d_schemes}. 
\subsection{Preliminaries of Combinatorial Design }\label{design}
Some basic useful definitions from combinatorial design are reviewed in this subsection. For further reading on this, readers can refer to \cite{Col,DPB,HM} and \cite{Stin}.

\begin{defn}[Design $(\mathcal{X}, \mathcal{A})$]\cite{Stin}
	A design is a pair  $(\mathcal{X}, \mathcal{A})$ such that the following properties are satisfied: \\
	\textit{1}. $\mathcal{X}$ is a set of elements called points, and\\
	\textit{2}. $\mathcal{A}$ is a collection (i.e., multiset) of nonempty subsets of $\mathcal{X}$ called blocks.
\end{defn}
\begin{defn}[Resolvable design]\cite{Stin}
A parallel class $\mathcal{P}$ in a design $(\mathcal{X}, \mathcal{A})$ is a subset of disjoint blocks from $\mathcal{A}$ whose union is $X$. A partition of $\mathcal{A}$ into $r$ parallel classes is called a resolution, and $(\mathcal{X}, \mathcal{A})$ is said to be a resolvable design if $\mathcal{A}$ has at least one resolution.
\end{defn}
Resolvable designs are discussed in \cite{Stin}, \cite{Col} and \cite{HRW}. The construction of resolvable designs from linear block codes is discussed in \cite{TaR}.

\begin{defn}[Cross Resolvable Design]\cite{DPB}
	For any resolvable design $(\mathcal{X}, \mathcal{A})$ with $r$ parallel classes, if there exist at least one $i\in \{2,3,\dots,r\}$ such that the cardinality of intersection of $i$ blocks drawn from any $i$ distinct parallel classes, denoted by $\mu_{i}$, exists and this value remains same ($\mu_i\neq 0$) for all possible choices of blocks, then the resolvable design is said to be a Cross Resolvable Design (CRD). A CRD for which $\mu_{r}$ exists is called a Maximal Cross Resolvable Design (MCRD).
\end{defn}
\begin{example}\label{ex:res_des2}
	$\mathcal{X}=\{1,2,3,4\}$, $ \mathcal{A}=\{12, 13, 14, 23, 24, 34\}$ is a resolvable design which can be partitioned into $3$ parallel classes  $\mathcal{P}_1=\{12, 34\}$, $\mathcal{P}_2=\{13, 24\}$ and $\mathcal{P}_3=\{14, 23\}$. This is a CRD with $\mu_{2}=1$. If $ \mathcal{A}=\{12, 13, 24, 34\}$, then it is an MCRD with $\mu_{2}=1$, where $\mathcal{P}_1=\{12, 34\}$ and $\mathcal{P}_2=\{13, 24\}$.
\end{example}
\begin{cor}\label{cor:CRD}
	For any positive integer $n \geq 2$, there exist an MCRD $(\mathcal{X}, \mathcal{A})$ with $\mathcal{X}=[n^2]$, $|\mathcal{A}|=2n$, $n$ blocks in each of the $2$ parallel classes and $\mu_{2}=1$.
\end{cor}
The proof of Corollary \ref{cor:CRD} is given in Appendix \textit{\ref{appendix:cor:CRD}}. 	
\begin{example}\label{eg:crd_cor}
	For $n=3$ by Corollary \ref{cor:CRD}, one can get an MCRD  $\mathcal{X}=\{1,2,3,4,5,6,7,8,9\}$, $\mathcal{A}=\{123,456,789,147,258,369\}$ with $2$ parallel classes, $\mathcal{P}_1=\{123, 456, 789\}$, $\mathcal{P}_2=\{147,258,369\}$ and $\mu_{2}=1$.  
\end{example}
\begin{defn}[$t$-GDD]\cite{HM}\label{def_tgdd}
	Let $v$ be a non-negative integer, $\lambda$ and $t$ be positive integers and $K$ be a set of positive integers.  A $t$-group divisible design ($t$-GDD) of order $v$, index $\lambda$ and block sizes from $K$ is a triple $(\mathcal{X}, \mathcal{G}, \mathcal{A})$ where \\
	\textit{1}. $\mathcal{X}$ is a set of $v$ elements called points, \\
	\textit{2}. $\mathcal{G}=\{G_1,G_2,...\}$ is a nonempty subsets of $\mathcal{X}$ which partition $\mathcal{X}$ called groups, \\
	\textit{3}. $\mathcal{A}$ is a family of subsets of $\mathcal{X}$ each of cardinality from $K$ called blocks such that each block intersects any given group in at most one point, \\
	\textit{4}. each $t$-set of points from $t$ distinct groups is in exactly $\lambda$ blocks.
\end{defn}
The $t$-GDDs with equal group sizes are called uniform. We denote a $t$-GDD with block size $k$ and $m$ groups of uniform group size $q$ ($\therefore$ $v=mq$) as $t-(m,q,k,\lambda)$ GDD. In a $t-(m,q,k,\lambda)$ GDD, the number of blocks in $\mathcal{A}$ is $\frac{\lambda\binom{m}{t}q^t}{\binom{k}{t}}$ \cite{SHMWT}.
\begin{example}\label{eg:GDD}
	$\mathcal{X}=\{1,2,3,4,5,6\}$, $\mathcal{G}=\{G_1=\{1,2\}, G_2=\{3,4\},G_3=\{5,6\}\}$ and $\mathcal{A}=\{\{1,3,5\}, \{2,3,6\},  \{1,4,6\}, \{2,4,5\}\}$ is a $2-(3,2,3,1)$ GDD.
\end{example}
\begin{thm}\label{thm:tgdd}\cite{SHMWT}
	Suppose that $(\mathcal{X}, \mathcal{G}, \mathcal{A})$ is a $t-(m,q,k,\lambda)$ GDD. Then, for any $i \in [t]$, a $t-(m,q,k,\lambda)$ GDD is also a $i-(m,q,k,\lambda_{i})$ GDD with $\lambda_{i}=\frac{\lambda q^{t-i} \binom{m-i}{t-i}}{\binom{k-i}{t-i}}$.
\end{thm}
From Theorem \ref{thm:tgdd} we can say that in a $t-(m,q,k,\lambda)$ GDD, for any $i \in [t]$, each $i$-set of points from $i$ distinct groups is in exactly $\lambda_{i}=\frac{\lambda q^{t-i} \binom{m-i}{t-i}}{\binom{k-i}{t-i}}$ blocks.
One can always construct a $2$-GDD for any positive integer $n>2$ as stated in the following corollary.
\begin{cor}\label{triv_GDD}
	For any positive integer $n>2$, there always exists a $2-(n,2,2,1)$ GDD.
\end{cor}
\begin{IEEEproof}
	Consider any positive integer $n>2$. Take  $\mathcal{X}=[2n]$, $\mathcal{G}=\{G_1,G_2,...,G_n\}$ where $G_i=\{2i-1, 2i\}$, $\forall i \in [n]$ and $\mathcal{A}$ is the collection of all $2$-set of points selected from $2$ distinct groups among the $n$ groups. That is, $\mathcal{A}=\{A_1,A_2,..,A_{2n(n-1)}\} :A_j=\{k, l\}$, $k \in G_i, l \in G_{i'}, i \ne i', \forall j \in [2n(n-1)]$. It is easy to see that $(\mathcal{X}, \mathcal{G}, \mathcal{A})$ is a $2-(n,2,2,1)$ GDD.
\end{IEEEproof}
\begin{defn}[Cyclic set]\cite{cyc_des}
	A cyclic set is a family of subsets (blocks) of $\mathbb{Z}_n$, the cyclic group of integers modulo $n$, where each subset is obtained by cyclically shifting a base subset $\mathcal{B} \subseteq \mathbb{Z}_n$. The cyclic set generated by $\mathcal{B}$ is $\{\mathcal{B} + i : i \in \mathbb{Z}_n\}$, where $\mathcal{B} + i = \{(b +i) \text{ mod } n : \forall b \in \mathcal{B} \}$. A full cyclic set is a cyclic set where all the cyclic shifts of the base block produce distinct subsets.
\end{defn}
\begin{example}
	Let $\mathcal{B}=\{0,1,3\}$ in $\mathbb{Z}_7$. The set $\{ \{0,1,3\},\{1,2,4\},\{2,3,5\},\{3,4,6\},\{4,5,0\},\{5,6,1\}, \{6,0,2\}\}$ is a full cyclic set.
\end{example}
\section{Lower bound on the transmission load of a DPDA}\label{lb_dpda}
In this section, we discuss the existing lower bound on the transmission load of a DPDA and propose a new lower bound on the transmission load of a DPDA. First, we define an \textit{optimal DPDA} that results in a D2D scheme that achieve a lower bound on the transmission load, as follows. 
\begin{defn}[Optimal DPDA]
	For a D2D coded caching problem with a given number of users $K$, number of files $N$, subpacketization level $F$, and each user having a cache of size $M$ files which can be represented as a $(K, F, Z, S)$ DPDA $\mathbf{P}$, where $Z=\frac{MF}{N}$, the DPDA is said to be an \textit{optimal DPDA} if the transmission load of the D2D coded caching scheme generated by the DPDA, $R=\frac{S}{F}$, is the least possible $R$ such that a $(K, F, Z, S)$ DPDA $\mathbf{P}$ exists. 
\end{defn}
For a given $(K,F,Z,S)$ DPDA, a lower bound on $R$ is given in \cite{JMQX}, as stated in the following theorem.
\begin{thm}\label{thm:JMQX}
	(\hspace{1sp}\cite{JMQX}) For any $(K,F,Z,S)$ DPDA $\mathbf{P}$, the transmission load of the D2D coded caching scheme generated by $\mathbf{P}$ satisfies  
	\begin{equation}\label{eq:JMQX}
		R=\frac{S}{F} \geq \left(\frac{F}{Z}-1\right),
	\end{equation} the equality holds if and only if every integer appears $\frac{KZ}{F}$ times and each row has exactly $\frac{KZ}{F}$ $'\star'$s. 	
\end{thm}
The JCM scheme achieves the lower bound in (\ref{eq:JMQX}) and therefore the corresponding DPDA is an optimal DPDA with respect to the lower bound in (\ref{eq:JMQX}). We propose a new lower bound on the transmission load $R$ of a DPDA as follows.
\begin{thm}\label{thm:lb_dpda}
	 For any $(K,F,Z,S)$ DPDA $\mathbf{P}$, the transmission load of the D2D coded caching scheme generated by $\mathbf{P}$ satisfies   
	\begin{equation}\label{eq:lb_dpda}
		R \geq \frac{K}{F}\left(\frac{F}{Z}-1\right),
	\end{equation}	 the equality holds if and only if every integer appears $Z$ times in $\mathbf{P}$.
\end{thm}
\begin{IEEEproof}
	For any $(K,F,Z,S)$ DPDA $\mathbf{P}$, by the condition \textit{C1} of DPDA definition, the total number of $'\star'$s in $\mathbf{P}$ is $KZ$ and non-$\star$ entries are $K(F-Z)$. By the condition \textit{C4} of the DPDA definition, each integer can appear at most $Z$ times. Therefore, $S \geq \frac{K(F-Z)}{Z} =K\left(\frac{F}{Z}-1\right)$. Thus, the transmission load $R=\frac{S}{F} \geq \frac{K}{F}\left(\frac{F}{Z}-1\right)$.
\end{IEEEproof}
	\begin{example}\label{eg:opt1}
		For $K=6,F=4,Z=2$, by  Theorem~\ref{thm:lb_dpda}, $S \geq 6$. Therefore, the $(6,4,2,6)$ DPDA given in Example \ref{eg:dpda} is an \textit{optimal DPDA} with respect to the lower bound in (\ref{eq:lb_dpda}).
	\end{example}
	When $K<F$, (\ref{eq:JMQX}) is tighter than (\ref{eq:lb_dpda}); when $K>F$, (\ref{eq:lb_dpda}) is tighter than (\ref{eq:JMQX}); and when $K=F$, both bounds are the same.
\begin{rem}
	From Table \ref{tab:d2d_schemes}, it can be verified that the JCM scheme and the schemes in Theorems $5$, $6$ and $7$ in \cite{JMQX_Arxiv} (Rows $3$ to $5$ of Table \ref{tab:d2d_schemes} ), satisfy the lower bound in (\ref{eq:JMQX}). Other than these, no other scheme satisfies either the lower bound in (\ref{eq:JMQX}) or that in (\ref{eq:lb_dpda}).
\end{rem}
\section{Proposed DPDA Constructions and Performance Analysis}\label{constr_d2d}	
In this section, we describe the construction of five classes of optimal DPDAs that give D2D coded caching schemes with low subpacketization levels using combinatorial designs. Theorem \ref{thm:d2d} only provides a D2D scheme from a given DPDA and does not describe a method for constructing DPDAs. Therefore, the construction of novel classes of DPDAs is both of theoretical interest and practical relevance. For each proposed DPDA construction, the performance analysis is carried out by comparing it with existing D2D coded caching schemes with the same system parameters, and by showing the optimality of the resulting DPDAs.
\subsection{Construction I}\label{constrn_dpda1}
In this subsection, we first construct a class of DPDAs by using the MCRD specified in Corollary \ref{cor:CRD}. This construction is provided in the proof of Theorem \ref{thm:dpda1}, which formally states the proposed class of DPDAs. We then generalize this construction to obtain DPDAs from any given CRD with $\mu_{2}=1$, as stated in Corollary \ref{cor:conc1_gen}. The idea is to represent users in a D2D network by the blocks and file packets by the points of a given CRD, and leverage the properties of CRDs to satisfy the conditions in the DPDA definition (Definition \ref{def_dpda}).  
 
\begin{thm}\label{thm:dpda1}
	For any positive integer $n \geq 2$, there exists a $\left( 2n,n^2,n, n^2(n-1) \right)$ DPDA obtainable from an MCRD with $\mu_{2}=1$, which gives a D2D coded caching scheme with memory ratio $\frac{M}{N}=\frac{1}{n}$ and transmission load $R=(n-1)$.
\end{thm}
\begin{IEEEproof}
For any positive integer $n$, using Corollary \ref{cor:CRD}, we obtain an MCRD $(\mathcal{X}, \mathcal{A})$ with $\mu_{2}=1$. We denote a block $A \in \mathcal{A}$ as $A_i^j : i \in [n], j \in [2]$, where $A_i^j$ denotes the $i^{th}$ block in $j^{th}$ parallel class $\mathcal{P}_j$. We use the notation $A^j(x)$ to denote the block $A$ in the $j^{th}$ parallel class to which point $x$ belongs. We construct an array $\mathbf{P}$  whose rows are indexed by all the points $x \in \mathcal{X}$ and columns are indexed by all the blocks $A_i^j \in \mathcal{A}$, defined as follows.   	
\begin{equation}\label{eq:dpda_constr1}
	P_{x,A_i^j}= \begin{cases}   \hspace{1cm}\star & \text {if } x \in A_i^j    \\  x \cup (A_i^j \cap A^{(j \mod 2)+1}(x)) & \text {otherwise. }\end{cases}
\end{equation}
The subpacketization level $F$ is the number of rows of $\mathbf{P}$, which is equal to $|\mathcal{X}| = n^2$. The number of users $K$ is given by the number of columns of $\mathbf{P}$, which is equal to   $|\mathcal{A}| = 2n$. The non-$\star$ entries of $\mathbf{P}$ are denoted by $x \cup (A_i^j \cap A^{(j \mod 2)+1}(x))$. Since $(\mathcal{X}, \mathcal{A})$ is an MCRD with $\mu_{2}=1$ and, $A_i^j$ and $A^{(j \mod 2)+1}(x)$ belongs to different parallel classes, $|(A_i^j \cap A^{(j \mod 2)+1}(x))|=1$. Therefore, $|x \cup (A_i^j \cap A^{(j \mod 2)+1}(x))|=2$ since $x \notin A_i^j$. Therefore, the number of possible $x \cup (A_i^j \cap A^{(j \mod 2)+1}(x))$ is the number of $2$ sized subsets of each of the $2n$ blocks. Since $\mu_{2}=1$, a $2$ sized subset of a given block of size $n$ does not appear in any other block. Therefore, $S= \binom{n}{2}2n=n^2(n-1)$. The mapping $\phi$ from the non-$\star$ entries $\{a,b\} \in \binom{A_i^j}{2}$  to the users, denoted by blocks $A_i^j \in \mathcal{A}$, is defined as follows.  $\phi(\{a,b\})= A_i^j \text{ if } \{a,b\} \subseteq A_i^j $. Now we have to show that the array $\mathbf{P}$ is the DPDA specified in Theorem \ref{thm:dpda1}, which is shown in Appendix \textit{\ref{appendix:dpda1}}. 
\end{IEEEproof}
The following example illustrates Theorem \ref{thm:dpda1}.

\begin{example}\label{ex:dpda_eg1}
	For $n=3$ we obtain the MCRD in Example \ref{eg:crd_cor}. The users are denoted by the six blocks in $\mathcal{A}$. Each file is divided into nine sub-packets, that is, $\{W_{n}=W_{n,x} : x \in [9], \forall n \in [N]\}$. Then, using (\ref{eq:dpda_constr1}), the $9 \times 6$ array shown in Fig.\ref{fig:dpda_eg1} is obtained. It can be verified that this array is a $(6,9,3,18)$ DPDA. From this DPDA, using Algorithm \ref{d2d_alg}, one can obtain a D2D coded caching scheme with $F=9$, $\frac{M}{N}=\frac{1}{3}$ and $R=2$. The mapping is $\phi(\{a,b\})= A_i^j \text{ if } \{a,b\} \subset A_i^j $. For example, consider the entry $12$. There is only one block $123$ in which $12$ appears. Since every $2$ sized non-$\star$ entry appears in exactly one block, there is a unique user corresponding to each non-$\star$ entry, and each user broadcasts $\binom{3}{2}=3$ messages. 
	\vspace{-0.1cm}
	\begin{figure}[!htbp]
	\centering\[	\footnotesize 
		\begin{array}{cc|*{7}{c}}
			& & \multicolumn{3}{c}{\text{$\mathcal{P}_1$}} & \vdots	& \multicolumn{3}{c}{\text{$\mathcal{P}_2$}} \\
			\hline
			A_i^j & & 123 & 456 & 789  & \vdots	& 147 & 258 & 369  \\
			\hline
			&	1 & \star & 14 & 17 &\vdots& \star & 12 & 13 \\	
			&	2 & \star & 25 & 28 &\vdots& 12 & \star & 23 \\	
			&	3 & \star & 36 & 39 &\vdots& 13 & 23 & \star \\	
			x &	4 & 14 & \star & 47 &\vdots& \star & 45 & 46 \\	
			&	5 & 25 & \star & 58 &\vdots& 45 & \star & 56 \\	
			&	6 & 36 & \star & 69 &\vdots& 46 & 56 & \star \\	
			&	7 & 17 & 47 & \star &\vdots& \star & 78 & 79 \\	
			&	8 & 28 & 58 & \star &\vdots& 78 & \star & 89 \\	
			&	9 & 39 & 69 & \star &\vdots& 79 & 89 & \star \\	
		\end{array}
		\]
	\caption{$(6,9,3,18)$ DPDA in Example \ref{ex:dpda_eg1} }
	\label{fig:dpda_eg1}
	\end{figure}
\end{example}

	The construction used to obtain the class of DPDAs in Theorem \ref{thm:dpda1} using the MCRD in Corollary \ref{cor:CRD} can be extended to any CRD with $\mu_{2}=1$, as stated in Corollary \ref{cor:conc1_gen}. However, optimality is achieved only when the CRD with $\mu_{2}=1$ is an MCRD. 
\begin{cor}\label{cor:conc1_gen}
	For any CRD $(\mathcal{X},\mathcal{A})$ with $|\mathcal{X}|$ = $v$, $|\mathcal{A}|$ = $b$ , block size $k$, $r=\frac{bk}{v}$ parallel classes and $\mu_{2}=1$, there exists a $\left( b,v,k,  \binom{k}{2}b \right)$ DPDA which gives a D2D coded caching scheme with memory ratio $\frac{M}{N}=\frac{k}{v}$ and load $R=\frac{\binom{k}{2}b}{v}$.  
\end{cor}
The proof of Corollary \ref{cor:conc1_gen} is given in Appendix \textit{\ref{appendix:dpda1_gen} }. 
The following example illustrates Corollary \ref{cor:conc1_gen}.
\begin{example}\label{eg:conc1_gen}
	From a $(4,2)$ linear code over $GF(3)$, using the procedure explained in \cite{TaR}, one can obtain a resolvable design with $\mathcal{X}=\{0,1,2,3,4,5,6,7,8\}$ and $\mathcal{A}=\{012, 345, 678, 036, 147, 258, 057, 138, 246, 048, 237, 156\}$ which can be partitioned into $4$ parallel classes  $\mathcal{P}_1=\{012, 345, 678\}$, $\mathcal{P}_2=\{036, 147, 258\}$, $\mathcal{P}_3=\{057, 138, 246\}$ and $\mathcal{P}_4=\{048, 237, 156\}$. In fact, this resolvable design is a CRD with $\mu_2=1$. From this CRD, we obtain the $(12,9,3,36)$ DPDA shown in Fig.\ref{fig:dpda_eg}. The mapping is $\phi(\{a,b\})= A_i^j \text{ if } \{a,b\} \subset A_i^j $. For example, consider the entry $03$. There is only one block $036$ in which $03$ appears. Since every $2$ sized non-$\star$ entry appears in exactly one block, there is a unique user corresponding to each non-$\star$ entry, and each user broadcasts $\binom{3}{2}=3$ messages.
\end{example}
\begin{figure}[H]
	\footnotesize
	\setlength{\arraycolsep}{2pt}
	\centering\[
	\begin{array}{cc|*{15}{c}}
		& & \multicolumn{3}{c}{\text{$\mathcal{P}_1$}} & \vdots	& \multicolumn{3}{c}{\text{$\mathcal{P}_2$}} & \vdots	& \multicolumn{3}{c}{\text{$\mathcal{P}_3$}} & \vdots	& \multicolumn{3}{c}{\text{$\mathcal{P}_4$}} \\
		\hline
		A_i^j & & 012 & 345 & 678  & \vdots	& 036 & 147 & 258 & \vdots	& 057 & 138 & 246 & \vdots	& 048 & 237 & 156\\
		\hline
		&	0 & \star & 03 & 06 &\vdots& \star & 07 & 05 &\vdots& \star & 08 & 04 &\vdots& \star & 02 & 01\\	
		&	1 & \star & 14 & 17 &\vdots& 13 & \star & 18 &\vdots& 15 & \star & 16 &\vdots&  01 & 12 & \star\\	
		&	2 & \star & 25 & 28 &\vdots& 26 & 24 & \star &\vdots& 27 & 23 & \star &\vdots& 02 &\star & 12\\	
		x &	3 & 03 & \star & 36 &\vdots& \star & 45 & 46 &\vdots& 23 & \star & 46 &\vdots& 56 & \star & 13 \\	
		&	4 & 14 & \star & 47 &\vdots& 46 & \star & 24 &\vdots& 04 & 48 & \star &\vdots& \star & 34 & 45\\	
		&	5 & 25 & \star & 58 &\vdots& 05 & 57 & \star &\vdots& \star & 15 & 56 &\vdots& 45 & 35 & \star \\	
		&	6 & 06 & 36 & \star &\vdots& \star & 46 & 26 &\vdots& 56 & 16 & \star &\vdots& 68 & 67 & \star \\	
		&	7 & 17 & 47 & \star &\vdots& 07 & \star & 57 &\vdots& \star & 37 & 27 &\vdots& 78 & \star & 67\\	
		&	8 & 28 & 58 & \star &\vdots& 38 & 18 & \star &\vdots& 08 & \star & 48 &\vdots& \star & 78 & 68\\	
	\end{array}
	\]
	\caption{$(12,9,3,36)$ DPDA in Example \ref{eg:conc1_gen} }
	\label{fig:dpda_eg}
\end{figure}
\begin{rem}
	The DPDA construction in Theorem $5$ of \cite{JMQX_Arxiv} has the same parameters as the DPDA in Theorem\ref{thm:dpda1}, but it is not obtained using CRDs. 
\end{rem}

The performance analysis of the DPDA in Theorem \ref{thm:dpda1} is carried out next. First, we show the optimality of the DPDA in Theorem \ref{thm:dpda1}. We then demonstrate its advantage in subpacketization over the JCM scheme, while achieving the same transmission load. Then, we present example-based comparisons: between the proposed scheme in Theorem \ref{thm:dpda1} and the hypercube scheme, and between the scheme in Corollary \ref{cor:conc1_gen} and the one derived from the coded caching scheme in \cite{TaR}, since a general comparison is not feasible.

\begin{thm}
	The DPDA obtained in Theorem \ref{thm:dpda1} is an \textit{optimal DPDA} with respect to the lower bound in (\ref{eq:JMQX}).
\end{thm}
\begin{IEEEproof}
	The transmission load of the D2D coded caching scheme generated by the  $\left( 2n,n^2,n, n^2(n-1) \right)$ DPDA in Theorem \ref{thm:dpda1} is $R=(n-1)$. By Theorem \ref{thm:JMQX}, $R=\frac{S}{F} \geq \left(\frac{F}{Z}-1\right) =\left(\frac{n^2}{n}-1\right)=(n-1)$. Therefore, the obtained DPDA is an optimal DPDA with respect to the lower bound in (\ref{eq:JMQX}).
\end{IEEEproof}
The JCM scheme satisfies the lower bound in (\ref{eq:JMQX}) and requires a subpacketization level that grows exponentially with the number of users. For $K=2n, \text{ }\forall n \geq 2 $ and $t=2$, the DPDA corresponding to the JCM scheme is a $\left( 2n,2n(2n-1),2(2n-1),2n(2n-1)(n-1) \right)$ DPDA which gives a D2D coded caching scheme with $\frac{M}{N}=\frac{1}{n}$ and transmission load $R=(n-1)$. That is, for the same number of users $K=2n \text{ }\forall n \geq 2$ and memory ratio $\frac{M}{N}=\frac{1}{n}$, both schemes achieve the same minimal transmission load $R=(n-1)$, but the subpacketization level of the proposed scheme is lower than that of the JCM scheme. That is, $\frac{F_{\text{proposed}}}{F_{\text{JCM scheme}}}=\frac{n}{2(2n-1)}$.
\begin{example}
	For $n=25$, the proposed construction gives a D2D scheme with $K=50$, $\frac{M}{N}=\frac{1}{25}$, $R=24$ and $F=625$, whereas for the same $K=50$ and $\frac{M}{N}=\frac{1}{25}$, the JCM scheme yields $R=24$ with $F=2450$. 
\end{example}
\begin{rem}
	Compared with the hypercube scheme in \cite{WCJ_HY}, both schemes give the same system parameters and $F$ for $n=2$, but the proposed scheme has improved $R$. For $n=2$, the proposed DPDA gives scheme for $K=4$ and $\frac{M}{N}=\frac{1}{2}$ with $F=4$ and load $R=1$,  where as the hypercube scheme provides a scheme for the same $K,\frac{M}{N}$ and $F$, but with $R=2$. 
\end{rem}
\begin{rem}
	Given a $(g+1)-(K,F,Z,S)$ PDA, Wang \textit{et al.} in \cite{JMQX} proposed a method to construct a $(K,gF,gZ,(g+1)S)$ DPDA. There exists no direct PDA construction using cross resolvable designs. In \cite{TaR}, low subpacketization coded caching schemes using resolvable designs were discussed. From a given $(n,k)$ linear code over $GF(q)$, the authors in \cite{TaR} constructed a resolvable design, and obtained a coded caching scheme from that. This can be represented as a PDA as follows. Given an $(n,k)$ linear code over $GF(q)$ that satisfies $(k,k+1)$ consecutive column property (refer to \cite{TaR} for details) generates a $(k+1)-\left( nq,q^k z,q^{k-1} z,\frac{q^k (q-1)nz}{k+1} \right)$ PDA, where $z$ is the least positive integer such that $(k+1)$ divides $nz$. From this PDA, using the construction in \cite{JMQX} one can obtain an $\left( nq,q^k kz,q^{k-1} kz,q^k (q-1)nz \right)$ DPDA.  In Example \ref{eg:conc1_gen}, from a $(4,2)$ linear code over $GF(3)$, a resolvable design with $4$ parallel classes is obtained. It can be verified that this design is a CRD with $\mu_{2}=1$. From this design, one can obtain a $3-(12,27,9,72)$ PDA, and from this PDA, using the construction in \cite{JMQX}, one can obtain a $(12,54,18,216)$ DPDA. This DPDA provides a D2D scheme with $K=12, \frac{M}{N}=\frac{1}{3}, F=54 $ and $R=4$. From the same CRD, we obtained the $(12,9,3,36)$ DPDA given in Example \ref{eg:conc1_gen}, which gives a D2D scheme with $K=12, \frac{M}{N}=\frac{1}{3}, F=9 $ and $R=4$. That is, for the same set of system parameters and achievable transmission load, the proposed scheme has a lower subpacketization level.
\end{rem}
\subsection{Construction II}\label{constrn_dpda2}
In this subsection, we construct a class of DPDAs using the $2-(n,2,2,1)$ GDD specified in Corollary \ref{triv_GDD}. This construction is provided in the proof of Theorem \ref{thm:dpda4}, which formally states the proposed class of DPDAs. The idea is to represent users in a D2D network by points and file packets by blocks of the $2-(n,2,2,1)$ GDD. The group divisibility property (property $2$ in Definition \ref{def_tgdd}) and the balancing property (property $4$ in Definition \ref{def_tgdd}) of a GDD help satisfy the conditions of the DPDA definition (Definition \ref{def_dpda}). 
\begin{thm}\label{thm:dpda4}
	For any positive integer $n > 2$, there exists a $( 2n,2n(n-1),2(n-1)^2, 2n )$ DPDA obtainable from a $2$-GDD, which gives a D2D coded caching scheme with memory ratio $\frac{M}{N}=1-\frac{1}{n}$ and transmission load $R=\frac{1}{n-1}$.
\end{thm}
\begin{IEEEproof}
	For any positive integer $n$, by Corollary \ref{triv_GDD}, we obtain a $2-(n,2,2,1)$ GDD $(\mathcal{X}, \mathcal{G}, \mathcal{A})$. We construct an array $\mathbf{P}$  whose rows are indexed by all blocks $A \in \mathcal{A}$ and columns are indexed by all points $x \in \mathcal{X}$, defined as follows.   	
	\begin{equation}\label{eq:dpda_constr4}
		P_{A,x}= \begin{cases}   \hspace{0.5cm}\star & \text {if } x \notin A   \\  A \backslash \{x\} & \text {otherwise. }\end{cases}
	\end{equation}
	The subpacketization level $F$ is the number of rows of $\mathbf{P}$, which is equal to $|\mathcal{A}| = 2n(n-1)$. The number of users $K$ is given by the number of columns of $\mathbf{P}$, which is equal to   $|\mathcal{X}| = 2n$. The non-$\star$ entries of $\mathbf{P}$ are denoted by $A \backslash \{x\}$. Since $|A|=2$, we have $|A \backslash \{x\}|=1$. Therefore, the number of possible $A \backslash \{x\}$ is the number of points in $\mathcal{X}$. Therefore, $S= 2n$. The mapping $\phi$ from the non-$\star$ entries $A\backslash \{x\}=\{y\}$  to the users, denoted by $\{x\} \in \mathcal{X}$, is defined as follows.  $\phi(\{y\})= \{x\} : x,y \in G_i, \text{for some } G_i \in \mathcal{G} $. Now we have to show that the array $\mathbf{P}$ is the DPDA specified in Theorem \ref{thm:dpda4}, which is shown in Appendix \textit{ \ref{appendix:dpda4}}.
\end{IEEEproof}
The following example illustrates Theorem \ref{thm:dpda4}.
\begin{example}\label{ex:dpda4}
	For $n=4$, Corollary \ref{triv_GDD} gives a $2-(4,2,2,1)$ GDD $(\mathcal{X}, \mathcal{G}, \mathcal{A})$ with $\mathcal{X}=[8]$, $\mathcal{G}=\{G_1=\{1,2\},G_2=\{3,4\},G_3=\{5,6\},G_4=\{7,8\}\}$ and $\mathcal{A} = \left\{ \{1,3\}, \{1,4\}, \{1,5\}, \{1,6\}, \{1,7\}, \{1,8\}, \{2,3\}, \{2,4\}, \right. \\ \left. \{2,5\},  \{2,6\}, \{2,7\}, \{2,8\}, \{3,5\}, \{3,6\}, \{3,7\}, \{3,8\},\right. \\ \left. \{4,5\},  \{4,6\}, \{4,7\}, \{4,8\}, \{5,7\}, \{5,8\}, \{6,7\}, \{6,8\} \right\}$. The users are denoted by all points in $\mathcal{X}$. Each file is divided into $24$ sub-packets denoted by blocks $A \in \mathcal{A}$ , that is, $\{W_{n}=W_{n,A} : A \in \mathcal{A}, \forall n \in [N]\}$. Then, using (\ref{eq:dpda_constr4}), we obtain the $24 \times 8$ array shown in Fig.\ref{fig:dpda4}. It can be verified that this array is a $(8,24,18,8)$ DPDA. From this DPDA, using Algorithm \ref{d2d_alg}, one can obtain a D2D coded caching scheme with $K=8$, $F=24$,  $\frac{M}{N}=\frac{3}{4}$ and $R=\frac{1}{3}$. The mapping is $\phi(\{y\})= \{x\} : x,y \in G_i, \text{for some } i \in [4] $.  For example, for the non-$\star$ entry $5$, we have $\phi(5)= 6$ since $5,6 \in G_3$.  	
\end{example}
\begin{figure}[H]
	\setlength{\arraycolsep}{3.5pt}		
	\footnotesize
	\centering \[
	\begin{array}{cc|cccccccc}
		x & & 1 & 2 & 3 & 4 & 5 & 6 & 7 & 8 \\
		\hline
		& \{1,3\} & 3 & \star & 1 & \star & \star & \star & \star & \star \\
		& \{1,4\} & 4 & \star & \star & 1 & \star & \star & \star & \star \\
		& \{1,5\} & 5 & \star & \star & \star & 1 & \star & \star & \star \\
		& \{1,6\} & 6 & \star & \star & \star & \star & 1 & \star & \star \\
		& \{1,7\} & 7 & \star & \star & \star & \star & \star & 1 & \star \\
		& \{1,8\} & 8 & \star & \star & \star & \star & \star & \star & 1 \\
		& \{2,3\} & \star & 3 & 2 & \star & \star & \star & \star & \star \\
		& \{2,4\} & \star & 4 & \star & 2 & \star & \star & \star & \star \\
		& \{2,5\} & \star & 5 & \star & \star & 2 & \star & \star & \star \\
		& \{2,6\} & \star & 6 & \star & \star & \star & 2 & \star & \star \\
		A & \{2,7\} & \star & 7 & \star & \star & \star & \star & 2 & \star \\
		& \{2,8\} & \star & 8 & \star & \star & \star & \star & \star & 2 \\
		& \{3,5\} & \star & \star & 5 & \star & 3 & \star & \star & \star \\
		& \{3,6\} & \star & \star & 6 & \star & \star & 3 & \star & \star \\
		& \{3,7\} & \star & \star & 7 & \star & \star & \star & 3 & \star \\
		& \{3,8\} & \star & \star & 8 & \star & \star & \star & \star & 3 \\
		& \{4,5\} & \star & \star & \star & 5 & 4 & \star & \star & \star \\
		& \{4,6\} & \star & \star & \star & 6 & \star & 4 & \star & \star \\
		& \{4,7\} & \star & \star & \star & 7 & \star & \star & 4 & \star \\
		& \{4,8\} & \star & \star & \star & 8 & \star & \star & \star & 4 \\
		& \{5,7\} & \star & \star & \star & \star & 7 & \star & 5 & \star \\
		& \{5,8\} & \star & \star & \star & \star & 8  & \star & \star & 5\\
		& \{6,7\} & \star & \star & \star & \star & \star & 7 & 6 & \star \\
		& \{6,8\} & \star & \star & \star & \star & \star & 8 & \star & 6 \\
	\end{array}
	\]	
	\caption{$(8,24,18,8)$ DPDA in Example \ref{ex:dpda4} }
	\label{fig:dpda4}
\end{figure}
\begin{rem}
	The DPDA construction in Theorem $6$ of \cite{JMQX_Arxiv} has the same parameters as those in Theorem \ref{thm:dpda4}. However, it was constructed recursively, beginning with a base DPDA for $n=4$, and extending to larger values of $n$ in increments of $2$ via a specified recursive rule. In contrast, by using the $2-(n,2,2,1)$ GDD, we obtained a simplified direct construction that achieved the same parameters.
\end{rem}

The performance analysis of the DPDA in Theorem \ref{thm:dpda4} is carried out next. We first show the optimality of the DPDA in Theorem \ref{thm:dpda4} and then its subpacketization advantage over the JCM scheme while achieving the same transmission load.
\begin{thm}\label{thm:lb_dpda4}
	The DPDA obtained in Theorem \ref{thm:dpda4} is an \textit{optimal DPDA} with respect to the lower bound in (\ref{eq:JMQX}).
\end{thm}
\begin{IEEEproof}
	The transmission load of the D2D coded caching scheme generated by the  $\left( 2n,2n(n-1),2(n-1)^2, 2n \right)$ DPDA in Theorem \ref{thm:dpda4} is $R=\frac{1}{n-1}$. By Theorem \ref{thm:JMQX}, $R=\frac{S}{F} \geq \left(\frac{F}{Z}-1\right) =\left(\frac{2n(n-1)}{2(n-1)^2}-1\right)=\frac{1}{n-1}$. Therefore, the obtained DPDA is an optimal DPDA with respect to the lower bound in (\ref{eq:JMQX}). 
\end{IEEEproof}
For $K=2n$ and $t=2(n-1)$, the JCM scheme has the same system parameters and transmission load as the scheme in Theorem \ref{thm:dpda4}, but requires a higher subpacketization level of $F=2(n-1)\binom{2n}{2(n-1)}=2n(n-1)(2n-1)$. That is, $\frac{F_{\text{proposed }}}{F_{\text{JCM scheme}}}=\frac{1}{2n-1}$.
\subsection{Construction III}\label{sec_constrn3}
In this subsection, we construct a class of DPDAs using a combinatorial design that is obtainable from any given integer $n \ge 3$, as described below. For any integer $n \ge 3$, consider an ordered $n$-tuple $(1,2,3,...,n)$. This ordered $n$-tuple cyclically contains the $n$ ordered pairs \{$(1,2)$, $(2,3)$, $....$, $(n-1,n)$, $(n,1)$\}, and no others. Let $\mathcal{S}_n$ denote the set of these $n$ cyclically contained ordered pairs for a given $n$. Let $\mathcal{S}_n'=\{(x,y) : x,y \in [n], x \ne y \text{ and } (x,y) \notin \mathcal{S}_n \}$. Consider the combinatorial design $([n], \mathcal{S}_n')$. From this combinatorial design, we construct the class of DPDAs stated in Theorem \ref{thm:dpda5}. Users in a D2D network are represented by points and file packets by blocks of the design $([n], \mathcal{S}_n')$. The corresponding DPDA construction is provided in the proof of Theorem \ref{thm:dpda5}.
\begin{thm}\label{thm:dpda5}
	For any integer $n \ge 3$, there exists an $\left( n,n(n-2),(n-2)^2, 2n \right)$ DPDA, which gives a D2D coded caching scheme with memory ratio $\frac{M}{N}=1-\frac{2}{n}$ and transmission load $R=\frac{2}{n-2}$.
\end{thm}
\begin{IEEEproof}
	For any integer $n \ge 3$, consider the design $([n], \mathcal{S}_n')$ described above. We construct an array $\mathbf{P}$  whose rows are indexed by all ordered pairs in $\mathcal{S}_n'$, and columns are indexed by all integers $z \in [n]$, defined as follows.   	
	\begin{equation}\label{eq:dpda_constr5}
		P_{(x,y),z}= \begin{cases}   \hspace{0.5cm}\star & \text {if } z \notin \{x,y\}  \\ (\{x,y\} \backslash z)_\alpha & \text {otherwise, }\end{cases}
	\end{equation}
	where $\alpha =1 \text { if } z =x \text{ and } \alpha =2 \text { if } z =y$.
	
	The subpacketization level $F$ is the number of rows of $\mathbf{P}$, which is equal to $|\mathcal{S}_n'| = n(n-1)-n=n(n-2)$. The number of users $K$ is given by the number of columns of $\mathbf{P}$, which is equal to $n$. The non-$\star$ entries of $\mathbf{P}$ are denoted by $(\{x,y\} \backslash z)_\alpha$. Since $|(\{x,y\} \backslash z)|=1$, the number of possible $(\{x,y\} \backslash z)$ is $n$. The subscript $\alpha \in [2]$. Therefore, $S= 2n$. The mapping $\phi$ from the non-$\star$ entries $p_\alpha$  to the users, denoted by $p$, where $p \in (1,2,...,n)$, is defined as follows.  
	\begin{equation*}
		\phi\left(p_\alpha\right)= \begin{cases}    ((p-2) \text{ mod } n) + 1 & \text { if } \alpha = 1   \\ (p \text{ mod } n ) + 1 & \text { if } \alpha = 2. \end{cases}
	\end{equation*}
	That is, when $\alpha=1$ and $\alpha=2$, $\phi\left(p_\alpha\right)$ is the nearest left and nearest right entries of $p$ in cyclic fashion in the ordered $n$-tuple $(1,2,...,n)$, respectively, $\forall p \in [n]$. Now, we have to show that the array $\mathbf{P}$ is the DPDA specified in Theorem \ref{thm:dpda5}, which is shown in Appendix \textit{\ref{appendix:dpda5}}.
\end{IEEEproof}
\begin{example}\label{ex:dpda5}
	For $n=5$, consider the design $([5],\mathcal{S}'_5)$, where $\mathcal{S}'_5=\{(1,3),(1,4),(1,5),(2,4),(2,5),(2,1),(3,5),(3,1),(3,2),\\  (4,1),(4,2),(4,3),(5,2),(5,3),(5,4)\}$. Users are denoted by points in $[6]$. Each file is divided into $15$ sub-packets, that is, $\{W_{n}=W_{n,(x,y)} : (x,y) \in \mathcal{S}'_5, \forall n \in [N]\}$. Then, using (\ref{eq:dpda_constr5}), we obtained the $5 \times 15$ array shown in Fig.\ref{fig:dpda5}. It can be verified that the array shown in Fig.\ref{fig:dpda5} is a $(5,15,9,10)$ DPDA.
	\begin{figure}[!htbp]
		\setlength{\arraycolsep}{3pt}	
		\small
		\centering \[
		\begin{array}{cc|ccccc}
			x & & 1 & 2 & 3 & 4 & 5  \\
			\hline
			& (1,3) & 3_1 & \star & 1_2 & \star & \star \\
			& (1,4) & 4_1 & \star & \star & 1_2 & \star \\
			& (1,5) & 5_1 & \star & \star & \star & 1_2 \\
			& (2,4) & \star & 4_1 & \star & 2_2 & \star \\
			& (2,5) & \star & 5_1 & \star & \star & 2_2 \\
			& (2,1) & 2_2 & 1_1 & \star & \star & \star \\
			A & (3,5) & \star & \star & 5_1 & \star & 3_2 \\
			& (3,1) & 3_2 & \star & 1_1 & \star & \star \\
			& (3,2) & \star & 3_2 & 2_1 & \star & \star \\
			& (4,1) & 4_2 & \star & \star & 1_1 & \star \\
			& (4,2) & \star & 4_2 & \star & 2_1 & \star \\
			& (4,3) & \star & \star & 4_2 & 3_1 & \star \\
			& (5,2) & \star & 5_2 & \star & \star & 2_1 \\
			& (5,3) & \star & \star & 5_2 & \star & 3_1 \\
			& (5,4) & \star & \star & \star & 5_2 & 4_1 \\
		\end{array}
		\]	
		\caption{$(5,15,9,10)$ DPDA in Example \ref{ex:dpda5} }
		\label{fig:dpda5}
	\end{figure}
	
	 From this DPDA, using Algorithm \ref{d2d_alg}, one can obtain a D2D scheme with $K=5$, $F=15$,  $\frac{M}{N}=\frac{3}{5}$ and $R=\frac{2}{3}$. The mapping is $$ \phi\left(p_\alpha\right)= \begin{cases}    ((p-2) \text{ mod } 5) + 1 & \text { if } \alpha = 1   \\ (p \text{ mod } 5 ) + 1 & \text { if } \alpha = 2. \end{cases} $$ That is, $\phi(2_1)=\phi(5_2)=1, \phi(1_2)=\phi(3_1)=2, \phi(2_2)=\phi(4_1)=3, \phi(3_2)=\phi(5_1)=4, \text{ and } \phi(1_1)=\phi(4_2)=5$. 	
\end{example}

\begin{rem}
	The DPDA construction in Theorem $7$ of \cite{JMQX_Arxiv} has the same parameters as the DPDA in Theorem \ref{thm:dpda5} for odd values of $n$. However, it was constructed recursively, beginning with a base DPDA for $n=3$, and extending to larger values of $n$ in increments of $2$ via a specified recursive rule. In contrast, by using a combinatorial design described in the proof of Theorem \ref{thm:dpda5}, we obtained a simplified direct construction that achieved the same parameters.
\end{rem}

The performance analysis of the DPDA in Theorem \ref{thm:dpda5} is presented next. We first show that the DPDA in Theorem \ref{thm:dpda5} is optimal and then demonstrate its subpacketization advantage over the JCM scheme.
\begin{thm}\label{thm:lb_dpda5}
	The DPDA obtained in Theorem \ref{thm:dpda5} is an \textit{optimal DPDA} with respect to the lower bound in (\ref{eq:JMQX}).
\end{thm}
\begin{IEEEproof}
	The transmission load of the D2D coded caching scheme generated by the  $\left( n,n(n-2),(n-2)^2, 2n \right)$ DPDA in Theorem \ref{thm:dpda5} is $R=\frac{2}{n-2}$. By Theorem \ref{thm:JMQX}, $R=\frac{S}{F} \geq \left(\frac{F}{Z}-1\right) =\left(\frac{n(n-2)}{(n-2)^2}-1\right)=\frac{2}{n-2}$. Therefore, the obtained DPDA is an optimal DPDA with respect to the lower bound in (\ref{eq:JMQX}).
\end{IEEEproof}
For $K=n$ and $t=n-2$, the JCM scheme has the same system parameters and transmission load as the scheme in Theorem \ref{thm:dpda5}, but requires a higher subpacketization level of $F=(n-2)\binom{n}{n-2}=\frac{n(n-1)(n-2)}{2}$. That is, $\frac{F_{\text{proposed }}}{F_{\text{JCM scheme}}}=\frac{2}{n-1}$.
\subsection{Construction IV}\label{constrn_dpda4}
In this subsection, we construct a novel class of DPDAs using the MCRD specified in Corollary \ref{cor:CRD}, which is also used to obtain the class of DPDAs in Construction I. The idea is to represent users in a D2D network by points and file packets by blocks of the MCRD specified in Corollary \ref{cor:CRD}, whereas in Construction I, the users in a D2D network are represented by the blocks and file packets by the points of the same MCRD. This construction is provided in the proof of Theorem \ref{thm:dpda2}, which formally states the proposed class of DPDAs.

\begin{thm}\label{thm:dpda2}
	For any positive integer $n \geq 2$, there exists a $\left(n^2,2n, 2, n^2(n-1) \right)$ DPDA obtainable from an MCRD with $\mu_{2}=1$, which gives a D2D coded caching scheme with memory ratio $\frac{M}{N}=\frac{1}{n}$ and transmission load $R=\frac{n(n-1)}{2}$.
\end{thm}
\begin{IEEEproof}
	For any positive integer $n$, using Corollary \ref{cor:CRD}, we obtain an MCRD $(\mathcal{X}, \mathcal{A})$ with $\mu_{2}=1$. We follow the notations used in DPDA Construction I here also. We construct an array $\mathbf{P}$  whose rows are indexed by all the blocks $A_i^j \in \mathcal{A}$ and columns are indexed by all the points $x \in \mathcal{X}$, defined as follows. 
	\begin{equation}\label{eq:dpda_constr2}
	P_{A_i^j,x}= \begin{cases}   \hspace{1cm}\star & \text {if } x \in A_i^j    \\  (A_i^j \cap A^{(j \mod 2)+1}(x))_{\alpha} & \text {otherwise, }\end{cases}
\end{equation}	
	where the subscript $\alpha$ denotes the $\alpha^{th}$ occurrence of $(A_i^j \cap A^{(j \mod 2)+1}(x))$ from left to right in the row indexed by $A_i^j$.
	
	The subpacketization level $F$ is the number of rows of $\mathbf{P}$, which is equal to $|\mathcal{A}| = 2n$. The number of users $K$ is given by the number of columns of $\mathbf{P}$, which is equal to $|\mathcal{X}| = n^2$. The non-$\star$ entries of $\mathbf{P}$ are denoted by $(A_i^j \cap A^{(j \mod 2)+1}(x))_{\alpha}$. Since $(\mathcal{X}, \mathcal{A})$ is an MCRD with $\mu_{2}=1$ and, $A_i^j$ and $A^{(j \mod 2)+1}(x)$ belongs to different parallel classes, $|(A_i^j \cap A^{(j \mod 2)+1}(x))|=1$. Therefore, the number of possible $(A_i^j \cap A^{(j \mod 2)+1}(x))$ is the number of points in $\mathcal{X}$, which is $n^2$. Since $A_i^j \cap A^{(j \mod 2)+1}(x) \subseteq A_i^j$ and $|(A_i^j \cap A^{(j \mod 2)+1}(x))|=1$, $A_i^j \cap A^{(j \mod 2)+1}(x)$ is a point in $A_i^j$. By (\ref{eq:dpda_constr2}), the non-$\star$ entries in a given row $A_i^j$ are in those columns denoted by points that are not in $A_i^j$. Therefore in a given row $A_i^j$, a specific point $y \in A_i^j$ repeats at those columns denoted by the $(n-1)$ points other than $y$ in $A^{(j \mod 2)+1}(y)$. Therefore, $\alpha \in [(n-1)]$ and hence $S=n^2(n-1)$. The mapping $\phi$ from the non-$\star$ entries $\{ a_\alpha \} : a \in \mathcal{X}$  to the users, denoted by the points in $\mathcal{X}$, is defined as an identity function.  That is, $\phi(\{a_\alpha \})= a $. Now we have to show that the array $\mathbf{P}$ is the DPDA specified in Theorem \ref{thm:dpda2}, which is shown in Appendix \textit{ \ref{appendix:dpda2}}.
\end{IEEEproof}
The following examples illustrates Theorem \ref{thm:dpda2}.
\begin{example}\label{ex:dpda2_eg1}
	For $n=3$ we obtain the MCRD in Example \ref{eg:crd_cor}. The users are denoted by nine points in $\mathcal{X}$. Each file is divided into six sub-packets, that is, $\{W_{n}=W_{n,A_i^j} : A_i^j \in \mathcal{A}, \forall n \in [N]\}$. Then, using (\ref{eq:dpda_constr2}), we obtain the $6 \times 9$ array shown in Fig.\ref{fig:dpda2_eg1}. It can be verified that the array shown in Fig.\ref{fig:dpda2_eg1} is a $(9,6,2,18)$ DPDA. From this DPDA, using Algorithm \ref{d2d_alg}, one can obtain a D2D coded caching scheme with $F=6$, $\frac{M}{N}=\frac{1}{3}$ and $R=3$. The mapping is $\phi(\{a_\alpha \})= a $. As $\alpha=2$, each user broadcasts two coded messages.
\end{example}
	\begin{figure}[H]
		\small
		\setlength{\arraycolsep}{3.5pt}	
	\centering\[	
	\begin{array}{c|cc|*{9}{c}}
		& x & & 1 & 2 & 3  & 4 & 5 & 6 & 7 & 8 & 9  \\
		\hline
		\multirow{3}{*}{\text{$\mathcal{P}_1$}}&&	123 & \star & \star & \star & 1_1 & 2_1 & 3_1 & 1_2 & 2_2 & 3_2 \\	
		&&	456 & 4_1 & 5_1 & 6_1 & \star & \star & \star & 4_2 & 5_2 & 6_2 \\	
		&&	789 & 7_1 & 8_1 & 9_1 & 7_2 & 8_2 & 9_2 & \star & \star & \star \\
		\hdots& & \hdots& \hdotsfor{9}	\\
		\multirow{3}{*}{\text{$\mathcal{P}_2$}}&A_i^j	&	147 & \star & 1_1 & 1_2 & \star & 4_1 & 4_2 & \star & 7_1 & 7_2 \\	
		&&	258 & 2_1 & \star & 2_2 & 5_1 & \star & 5_2 & 8_1 & \star & 8_2 \\	
		&&	369 & 3_1 & 3_2 & \star & 6_1 & 6_2 & \star & 9_1 & 9_2 & \star \\	
	\end{array}
	\]
	\caption{$(9,6,2,18)$ DPDA in Example \ref{ex:dpda2_eg1} }
	\label{fig:dpda2_eg1}
\end{figure}

 The performance analysis of the DPDA in Theorem \ref{thm:dpda2} is carried out next.
First, we show the optimality of the DPDA in Theorem \ref{thm:dpda2}. We then compare the proposed scheme in Theorem \ref{thm:dpda2} with the JCM and hypercube schemes, demonstrating its advantage in subpacketization levels.
\begin{thm}\label{thm:lb_dpda2}
	The DPDA obtained in Theorem \ref{thm:dpda2} is an \textit{optimal DPDA} with respect to the lower bound in (\ref{eq:lb_dpda}).
\end{thm}
\begin{IEEEproof}
	The transmission load of the D2D coded caching scheme generated by the  $\left(n^2,2n, 2, n^2(n-1) \right)$ DPDA in Theorem \ref{thm:dpda2} is $R=\frac{n(n-1)}{2}$. By Theorem \ref{thm:lb_dpda}, $R=\frac{S}{F} \geq \frac{K}{F}\left(\frac{F}{Z}-1\right) =\frac{n^2}{2n}\left(\frac{2n}{2}-1\right)=\frac{n(n-1)}{2}$. Therefore, the obtained DPDA is an optimal DPDA with respect to the lower bound in (\ref{eq:lb_dpda}).
\end{IEEEproof}
A comparison of the proposed scheme in Theorem \ref{thm:dpda2} with the JCM scheme and hypercube scheme is given in Table \ref{tab:schm2_comp}. 
\begin{table}[h]
	\scriptsize
	\centering
	\begin{tabular}{| c | c | c| c |c |}
		\hline
		\rule{0pt}{4ex}
		\makecell{Schemes and \\ Parameters} & \makecell{ $K$} & \makecell{ $\frac{M}{N}$} & \makecell{$F$} & \makecell{ $R$}  \\ [4pt]	
		\hline
		\rule{0pt}{3.5ex}
		\makecell{JCM scheme in \cite{Ji} \\ $n \in \mathbb{Z}^{+}, K =n^2, t =n$ } & $n^2$ & $\frac{1}{n}$ & $n\binom{n^2}{n}$ &  $n-1$ \\
		\hline
		\rule{0pt}{3.5ex}
		\makecell{Hypercube scheme in \cite{WCJ_HY} \\  $n \in \mathbb{Z}^{+}$ and $n \geq 2 $  } & $n^2$ & $\frac{1}{n}$  & $n^n$ & $n$ \\		
		\hline
		\rule{0pt}{3.5ex}
		\makecell{Proposed scheme in Theorem \ref{thm:dpda2} \\ $n \in \mathbb{Z}^{+}$ and $n \geq 2 $  } & $n^2$ & $\frac{1}{n}$  & $2n$ & $\frac{n(n-1)}{2}$ \\
		\hline
	\end{tabular}
	\caption{Comparison of D2D scheme in Theorem \ref{thm:dpda2}}		
	\label{tab:schm2_comp}
\end{table} 
For the same number of users $n^2$ and memory ratio $\frac{1}{n}$, the subpacketization level of the proposed scheme grows sub-linearly, whereas those of the JCM and hypercube schemes vary exponentially and polynomially, respectively, with respect to the number of users. However, in general, the reduction in subpacketization occurs at the expense of an increase in transmission load. For sub-linear subpacketization, the achieved load is the minimal load as shown in Theorem \ref{thm:lb_dpda2}.  For $n=2$, the proposed scheme has an advantage in terms of both subpacketization and load, as shown in Table \ref{tab:schm2_comp_2}. For $n=3$, the proposed DPDA provides a scheme for $K=9$ and $\frac{M}{N}=\frac{1}{3}$ with $F=6$ and load $R=3$,  whereas the hypercube scheme achieves the same load $R=3$  for the same $K$ and $\frac{M}{N}$, but with a higher subpacketization level $F=27$.
\begin{table}[h]
	\scriptsize
	\centering
	\begin{tabular}{| c | c | c| c |c |}
		\hline
		\rule{0pt}{4ex}
		\makecell{Schemes for $n=2$} & \makecell{ $K$} & \makecell{ $\frac{M}{N}$} & \makecell{$F$} & \makecell{ $R$}  \\ [4pt]	
		\hline
		\rule{0pt}{3.5ex}
		\makecell{JCM scheme in \cite{Ji}} & $4$ & $\frac{1}{2}$ & $12$ &  $1$ \\
		\hline
		\rule{0pt}{3.5ex}
		\makecell{Hypercube scheme in \cite{WCJ_HY} } & $4$ & $\frac{1}{2}$  & $4$ & $2$ \\		
		\hline
		\rule{0pt}{3.5ex}
		\makecell{Proposed scheme in Theorem \ref{thm:dpda2} } & $4$ & $\frac{1}{2}$  & $4$ & $1$ \\
		\hline
	\end{tabular}	
	\caption{Comparison of schemes in Table \ref{tab:schm2_comp} for $n=2$}	
	\label{tab:schm2_comp_2}
\end{table}
\subsection{Construction V}\label{constrn_dpda5}
In this subsection, we construct a class of DPDAs using a combinatorial design obtainable from any given positive integers $k$ and $n$ such that $k\mid n$, as described below. For any positive integers $k$ and $n$ such that $k\mid n$, consider a design $(\mathcal{X}, \mathcal{A})$ with $\mathcal{X}=[n]$ and $\mathcal{A}=\binom{[n]}{k}$. The idea is to represent users in a D2D network by blocks and file packets by points of this design. This construction is provided in the proof of Theorem \ref{thm:dpda3}, which formally states the proposed class of DPDAs. 
\begin{thm}\label{thm:dpda3}
	For any positive integers $k$ and $n$ such that $k >1,n >1$ and $k\mid n$, there exists a $\left(\binom{n}{k},n, k, \binom{n}{k}\left(\frac{n}{k}-1 \right) \right)$ DPDA, which gives a D2D coded caching scheme with memory ratio $\frac{M}{N}=\frac{k}{n}$ and transmission load $R=\binom{n}{k} \frac{n-k}{nk}$.
\end{thm}
\begin{IEEEproof}
	For any positive integers $k$ and $n$ such that $k\mid n$, consider the combinatorial design $(\mathcal{X}=[n], \mathcal{A}=\binom{[n]}{k})$. We construct an array $\mathbf{P}$  whose rows are indexed by the points $i \in \mathcal{X}$ and columns are indexed by the blocks $\mathcal{C} \in \mathcal{A}$, defined as follows. 
	\begin{equation}\label{eq:dpda_constr3}
		P_{i,\mathcal{C}}= \begin{cases}   \hspace{1cm}\star & \text {if } i \in \mathcal{C}    \\  (i \cup \mathcal{D}(i,\mathcal{C}))_{\alpha} & \text {otherwise, }\end{cases}
	\end{equation}	
	where $\{\mathcal{D}(i,\mathcal{C}) : \mathcal{D}(i,\mathcal{C}) \subset \mathcal{C}, |\mathcal{D}(i,\mathcal{C})|=k-1\}$ is obtained using Algorithm \ref{alg:D} and the subscript $\alpha$ denotes the $\alpha^{th}$ occurrence of $(i \cup \mathcal{D}(i,\mathcal{C}))$ from left to right in the row indexed by $i$.
\begin{algorithm}[h]
	\renewcommand{\thealgorithm}{2}
	\caption{Algorithm for choosing $\mathcal{D}(i,\mathcal{C})$ in (\ref{eq:dpda_constr3}) }
	\label{alg:D}
	\begin{algorithmic}[1]
		\For{\texttt{$i \in [n]$}}
		\State Let $\mathcal{U}^{(i)}=\{[n]\backslash {i}\}=\{u_0,u_1,...,u_{n-2}\}$ be a ordered set (say lexicographic order).
		\State Define the collection of sets $\mathcal{C}^{(i)}=\{\mathcal{C} \in \binom{[n]}{k} : i \notin \mathcal{C} \}=\binom{\mathcal{U}^{(i)}}{k}$. Let $\mathcal{C}^{(i)}=\{\mathcal{C}^{(i)}_{1},\mathcal{C}^{(i)}_{2},...,\mathcal{C}^{(i)}_{m} \}$ in lexicographic ordering, where $m=\binom{n-1}{k}$.
		\State Rearrange the integer elements in each set of $\mathcal{C}^{(i)}$ such that each integer in $\mathcal{U}^{(i)}$ appears equal number of times as the first element in the ordered $k$-tuples.\textsuperscript{*} Let $\mathcal{C}'^{(i)}=\{\mathcal{C}'^{(i)}_{1},\mathcal{C}'^{(i)}_{2},...,\mathcal{C}'^{(i)}_{m} \}$ be the obtained collection of ordered $k$-tuples. This can be obtained as follows.
		\Procedure{}{Obtaining $\mathcal{C}'^{(i)}$ from $\mathcal{C}^{(i)}$ }
		\For{\texttt{$j=0 \text{ to } n-2$}}
		\State \vspace{-4ex} \begin{multline*} \hspace{1cm} \mathcal{T}_j^{(1)}=\{u_{j \text {mod } (n-1)}, u_{(j+1) \text {mod } (n-1)},..., \\ u_{(j+k-1) \text {mod } (n-1)} \} \end{multline*} \vspace{-4ex}
		\EndFor 
		\State Let $\mathcal{T}^{(1)}=\{\mathcal{T}_j^{(1)}, \forall j=0 \text{ to } n-2\}$
		\For{\texttt{$l=2 \text{ to } \frac{m}{n-1}$}}
		\State Choose $\mathcal{T}_0^{(l)}$ such that $ \mathcal{T}_0^{(l)} \in \mathcal{C}^{(i)} \backslash \bigcup_{x=1}^{l-1} \mathcal{T}^{(x)}$ and $u_0 \in \mathcal{T}_0^{(l)}$.
		\State Rearrange the elements in $\mathcal{T}_0^{(l)}$ such that $u_0$ comes as the first element. Let $\mathcal{T}_0^{(l)}=\{u_0, u_{a_1}, u_{a_2},...,u_{a_{k-1}}\}$, where $a_i \in [n-2]$.
		\For{\texttt{$j=0 \text{ to } n-2$}}
		\State \vspace{-4ex} \begin{multline*} \hspace{1.3cm} \mathcal{T}_j^{(l)}=\{u_{j \text {mod } (n-1)}, u_{(j+a_1) \text {mod } (n-1)},...,\\ u_{(j+a_{k-1}) \text {mod } (n-1)} \} \end{multline*} \vspace{-4ex}
		\EndFor
		\State $\mathcal{T}^{(l)}=\{\mathcal{T}_j^{(l)}, \forall j=0 \text{ to } n-2\}$
		\EndFor 
		\State $\mathcal{C}'^{(i)}=\{\mathcal{C}'^{(i)}_{1},\mathcal{C}'^{(i)}_{2},...,\mathcal{C}'^{(i)}_{m} \}$, where $\mathcal{C}'^{(i)}_{p}=\mathcal{T}_j^{(l)} : \mathcal{T}_j^{(l)}=\mathcal{C}^{(i)}_{p} \text{ (upto permutation) for some $j$ and $l$ }  , \forall p \in [m]$.
		\EndProcedure 
		\For{\texttt{$\mathcal{C}'^{(i)}_{p} \in \mathcal{C}'^{(i)}$, $p \in [m]$}}
		\State  Let $\mathcal{C}'^{(i)}_{p}=\{j_1,j_2,...,j_k\}$. Then $\mathcal{D}(i,\mathcal{C}^{(i)}_{p})=\{j_2,...,j_k\}$. That is, $\mathcal{D}(i,\mathcal{C}^{(i)}_{p})$ is obtained by removing the first element of $\mathcal{C}'^{(i)}_{p}$.
		\EndFor 
		\EndFor        	
	\end{algorithmic}
	\vspace{1mm}
	\hrule
	\vspace{1mm}
	\noindent\textsuperscript{*}\small{This rearrangement is possible since $k \nmid (n-1)$ ($\because$ if $k \mid n$, then $k \nmid (n-1)$). Since $k$ and $n-1$ are relatively prime, the $\binom{n-1}{k}$ distinct sets in $\binom{[n-1]}{k}$ can be set out in $\frac{m}{n-1}$ \textit{full cyclic sets} each having $n-1$ set entries \cite{cyc_des}. Thus, each $j \in \mathcal{U}^{(i)}$ appears as the first element in exactly $\frac{m}{n-1}$ of the $\mathcal{C}'^{(i)}_{x}$s. }
\end{algorithm}

	The subpacketization level $F$ is the number of rows of $\mathbf{P}$, which is equal to $n$. The number of users $K$ is given by the number of columns of $\mathbf{P}$, which is equal to $\binom{n}{k}$. The non-$\star$ entries of $\mathbf{P}$ are denoted by $(i \cup \mathcal{D}(i,\mathcal{C}))_{\alpha}$. Since $i \notin \mathcal{C}$ and $\{\mathcal{D}(i,\mathcal{C}) : \mathcal{D}(i,\mathcal{C}) \subset \mathcal{C}, |\mathcal{D}(i,\mathcal{C})|=k-1\}$, implies $|i \cup \mathcal{D}(i,\mathcal{C})|=k$. Therefore, the number of possible $(i \cup \mathcal{D}(i,\mathcal{C}))$ is $\binom{n}{k}$. For a given row indexed by $i$, from Algorithm \ref{alg:D} it is clear that, the number of possible $\mathcal{D}(i,\mathcal{C})$ is $\binom{n-1}{k-1}$. Also, $\mathcal{D}(i,\mathcal{C})$ for a given $i$ is obtained by Algorithm \ref{alg:D} by removing the first elements from the ordered k-tuples, which are uniformly distributed. Therefore, each $\mathcal{D}(i,\mathcal{C})$ occurs exactly $\frac{\binom{n-1}{k}}{\binom{n-1}{k-1}}=\frac{n}{k}-1$ times in a given row indexed by $i$. Therefore,  $\alpha \in \left[ \frac{n}{k}-1 \right]$ and hence $S=\binom{n}{k}\left(\frac{n}{k}-1 \right)$. The mapping $\phi$ from the non-$\star$ entries $(i \cup \mathcal{D}(i,\mathcal{C}))_{\alpha}$  to the users, denoted by $\mathcal{C}$, is defined as $\phi((i \cup \mathcal{D}(i,\mathcal{C}))_{\alpha})= \mathcal{C}$, where $ (i \cup \mathcal{D}(i,\mathcal{C}))=\mathcal{C} $. Now we have to show that the array $\mathbf{P}$ is the DPDA specified in Theorem \ref{thm:dpda3}, which is shown in Appendix \textit{\ref{appendix:dpda3}}.
\end{IEEEproof}

The following example illustrates Theorem \ref{thm:dpda3} and Algorithm \ref{alg:D}.
\begin{example}\label{ex:dpda3}
For $n=6$ and $k=2$, consider the design $(\mathcal{X}=[6],\mathcal{A}=\binom{[6]}{2})$. The users are denoted by the blocks in $\mathcal{A}$. Each file is divided into six sub-packets, that is, $\{W_{n}=W_{n,i} : i \in \mathcal{X}, \forall n \in [N]\}$. Then by (\ref{eq:dpda_constr3}), we obtain the $6 \times 15$ array shown in Fig.\ref{fig:dpda3}. It can be verified that the array shown in Fig.\ref{fig:dpda3} is a $(15,6,2,30)$ DPDA. From this DPDA, using Algorithm \ref{d2d_alg}, one can obtain a D2D coded caching scheme with $K=15$, $F=6$,  $\frac{M}{N}=\frac{1}{3}$ and $R=5$. The mapping is $\phi(\{x,y\}_{\alpha})= \{x,y\}$, $\forall \{x,y\} \in \mathcal{A}$. 
\begin{figure*}[!htbp]
	\setlength{\arraycolsep}{3.5pt}	
	\centering \[
	\begin{array}{cc|*{15}{c}}
		\mathcal{C} & & 12 & 13&14&15&16&23&24&25&26&34&35&36&45&46&56  \\
		\hline
		&	1 & \star & \star&\star&\star&\star & 13_1&14_1&12_1&12_2 & 14_2&15_1&13_2&15_2&16_1&16_2 \\	
		&	2 & \star & 23_1&24_1&12_1&12_2 & \star&\star&\star&\star & 24_2&25_1&23_2&25_2&26_1&26_2 \\	
		&	3 & 23_1 & \star&34_1&13_1&13_2 & \star&34_2&35_1&23_2 & \star&\star&\star&35_2&36_1&36_2 \\
		i &	4 & 24_1 & 34_1&\star&14_1&14_2& 34_2&\star&45_1&24_2&\star & 45_2&46_1&\star&\star&46_2 \\
		&	5 & 25_1 & 35_1&15_1&\star&15_2 & 35_2&45_1&\star&25_2 & 45_2&\star&56_1&\star&56_2&\star \\	
		&	6 & 26_1 & 36_1&16_1&16_2&\star & 36_2&46_1&26_2&\star & 46_2&56_1&\star&56_2&\star&\star \\
	\end{array}
	\] 
	\caption{$(15,6,2,30)$ DPDA in Example \ref{ex:dpda3} }
	\label{fig:dpda3}
\end{figure*}

To clarify how the non-$\star$ entries are obtained, consider the row $i=1$. By Lines $2$ and $3$ of Algorithm \ref{alg:D}, $\mathcal{U}^{(i)}=\{ u_0=2,u_1=3,u_2=4,u_3=5,u_4=6\}$ and $\mathcal{C}^{(1)}=\{ \mathcal{C}^{(1)}_{1}=23,\mathcal{C}^{(1)}_{2}=24,\mathcal{C}^{(1)}_{3}=25,\mathcal{C}^{(1)}_{4}=26,\mathcal{C}^{(1)}_{5}=34,\mathcal{C}^{(1)}_{6}=35,\mathcal{C}^{(1)}_{7}=36,\mathcal{C}^{(1)}_{8}=45,\mathcal{C}^{(1)}_{9}=46,\mathcal{C}^{(1)}_{10}=56\}$.  To obtain $\mathcal{C}'^{(1)}$, we use the procedure explained in Lines $5-19$ of Algorithm \ref{alg:D}. By Lines $6-9$, we obtain $\mathcal{T}^{(1)}=\{ \mathcal{T}_0^{(1)}=\{2,3\}=23, \mathcal{T}_1^{(1)}=34, \mathcal{T}_2^{(1)}=45, \mathcal{T}_3^{(1)}=56, \mathcal{T}_4^{(1)}=62\}$. By Line $11$, choose $\mathcal{T}_0^{(2)}=\{u_0=2,u_{a_1}=4\}=24$. Then by lines $13-16$, we obtain $\mathcal{T}^{(2)}=\{ \mathcal{T}_0^{(2)}=24, \mathcal{T}_1^{(2)}=35, \mathcal{T}_2^{(2)}=46, \mathcal{T}_3^{(2)}=52, \mathcal{T}_4^{(2)}=63\}$. Thus, by Line $18$ we obtain $\mathcal{C}'^{(1)}=\{\mathcal{C}'^{(1)}_{1}=\mathcal{T}_0^{(1)}=23,\mathcal{C}'^{(1)}_{2}=\mathcal{T}_0^{(2)}=24,\mathcal{C}'^{(1)}_{3}=\mathcal{T}_3^{(2)}=52,\mathcal{C}'^{(1)}_{4}=\mathcal{T}_4^{(1)}=62,\mathcal{C}'^{(1)}_{5}=\mathcal{T}_1^{(1)}=34,\mathcal{C}'^{(1)}_{6}=\mathcal{T}_1^{(2)}=35,\mathcal{C}'^{(1)}_{7}=\mathcal{T}_4^{(2)}=63,\mathcal{C}'^{(1)}_{8}=\mathcal{T}_2^{(1)}=45,\mathcal{C}'^{(1)}_{9}=\mathcal{T}_2^{(2)}=46,\mathcal{C}'^{(1)}_{10}=\mathcal{T}_3^{(1)}=56 \}$.  By Line $21$ of Algorithm \ref{alg:D}, the corresponding $\mathcal{D}(1,\mathcal{C})$ entries, in the same order, are $\{3,4,2,2,4,5,3,5,6,6\}$. Thus, the corresponding $(i \cup \mathcal{D}(1,\mathcal{C}))$s are $\{13,14,12,12,14,15,13,15,16,16\}$. Each distinct entry appears $\frac{n}{k}-1=2$ times in this row. Therefore, $\alpha \in [2]$.  Hence, by (\ref{eq:dpda_constr3}), the non-$\star$ entries in the row $i=1$ are $\{13_1,14_1,12_1,12_2,14_2,15_1,13_2,15_2,16_1,16_2\}$, as evident from Fig.\ref{fig:dpda3}.
\end{example}
\begin{rem}
	It has to be noted that only the $\star$ entries in the transpose of DPDA in Construction \ref{constrn_dpda1} are the same as the $\star$ entries in the DPDA in Construction \ref{constrn_dpda4}. The transpose of DPDA in Construction \ref{constrn_dpda1}  is not a DPDA. Similarly, the $\star$ entries in the transpose of the MAN PDA in \cite{YCT} are the same as the $\star$ entries in the DPDA in Construction \ref{constrn_dpda5}. The transpose of the MAN PDA is a PDA, and not a DPDA. 
\end{rem}

The performance analysis of the DPDA in Theorem \ref{thm:dpda3} is presented next.
First, we show the optimality of the DPDA in Theorem \ref{thm:dpda3}. We then compare the proposed scheme with the JCM scheme and the scheme in Corollary $6$ of \cite{MADCC_arxiv}, demonstrating an advantage in subpacketization over the JCM scheme, and in both subpacketization and transmission load over the scheme in Corollary $6$ of \cite{MADCC_arxiv}.
\begin{thm}\label{thm:lb_dpda3}
	The DPDA obtained in Theorem \ref{thm:dpda3} is an \textit{optimal DPDA} with respect to the lower bound in (\ref{eq:lb_dpda}).
\end{thm}
\begin{IEEEproof}
	The transmission load of the D2D coded caching scheme generated by the  $\left(\binom{n}{k},n, k, \binom{n}{k}\left(\frac{n}{k}-1 \right) \right)$ DPDA in Theorem \ref{thm:dpda3} is $R=\binom{n}{k} \frac{n-k}{nk}$. By Theorem \ref{thm:lb_dpda}, $R=\frac{S}{F} \geq \frac{K}{F}\left(\frac{F}{Z}-1\right) =\frac{\binom{n}{k}}{n}\left(\frac{n}{k}-1\right)=\binom{n}{k} \frac{n-k}{nk}$. Therefore, the obtained DPDA is an optimal DPDA with respect to the lower bound in (\ref{eq:lb_dpda}).
\end{IEEEproof}
A comparison of the proposed scheme in Theorem \ref{thm:dpda3} with the JCM scheme and scheme in Corollary $6$ in \cite{MADCC_arxiv} (Row $14$ of Table \ref{tab:d2d_schemes}), for the same system parameters $K$ and $\frac{M}{N}$, is given in Table \ref{tab:schm5_comp}. 
\begin{table}[h]
	\scriptsize
	\centering
	\setlength{\tabcolsep}{3pt}
	\begin{tabular}{| c | c | c| c |c |}
		\hline
		\rule{0pt}{4ex}
		\makecell{Schemes and \\ Parameters} & \makecell{ $K$} & \makecell{ $\frac{M}{N}$} & \makecell{$F$} & \makecell{ $R$}  \\ [4pt]	
		\hline
		\rule{0pt}{3.5ex}
		\makecell{JCM scheme in \cite{Ji} \\ $n \in \mathbb{Z}^{+}, K =\binom{n}{k}, t =\binom{n-1}{k-1}$ } & $\binom{n}{k}$ & $\frac{k}{n}$ & $\binom{n-1}{k-1}\binom{\binom{n}{k}}{\binom{n-1}{k-1}}$ &  $\frac{n-k}{k}$ \\
		\hline
		\rule{0pt}{3.5ex}
		\makecell{Scheme in Corollary $6$ in \cite{MADCC_arxiv} \\  $n,k \in \mathbb{Z}^{+}$ and $k >1$,\\ $n >1$, $k\mid n$ and $i=1$  } & $\binom{n}{k}$ & $\frac{k}{n}$  & $n\binom{k}{2}$ & $\binom{n}{k}\frac{n-k}{n(k-1)}$ \\		
		\hline
		\rule{0pt}{3.5ex}
		\makecell{Proposed scheme in Theorem \ref{thm:dpda3} \\ $n,k \in \mathbb{Z}^{+}$ and $k >1$,\\ $n >1$ and $k\mid n$  } & $\binom{n}{k}$ & $\frac{k}{n}$  & $n$ & $\binom{n}{k}\frac{n-k}{nk}$ \\
		\hline
	\end{tabular}
	\caption{Comparison of D2D scheme in Theorem \ref{thm:dpda3}}		
	\label{tab:schm5_comp}
\end{table} 

From Table \ref{tab:schm5_comp}, it is clear that, compared to the JCM scheme, the proposed scheme in Theorem \ref{thm:dpda3} has a significantly lower subpacketization level at the expense of an increase in transmission load. Compared with the scheme in   Corollary $6$ in \cite{MADCC_arxiv}, the proposed scheme in Theorem \ref{thm:dpda3}  has an advantage in both the subpacketization level and transmission load. That is, $\frac{F_{\text{proposed }}}{F_{\text{Corollary 6 in [13]}}}=\frac{1}{\binom{k}{2}}$  and  $\frac{R_{\text{proposed }}}{R_{\text{Corollary 6 in [13]}}}=\frac{k-1}{k}$.
\begin{rem}
	A comparison of the proposed schemes with the four DPDA constructions in \cite{JY} is not possible because all of them deal with larger memories, and these memory points are not covered by our constructions.  
\end{rem}
\section{Conclusion}\label{concl_d2d}
In this work, we constructed five classes of DPDAs using combinatorial designs that provide low subpacketization level D2D schemes. The first three classes of constructed DPDAs achieve the known lower bound on the transmission load of DPDA while requiring a subpacketization level lower than that of the JCM scheme. We then proposed another lower bound on the transmission load of a DPDA and constructed two classes of novel DPDAs that achieve this lower bound. Constructing new classes of DPDAs that achieve the lower bound on the transmission load of a DPDA is an interesting open problem.

\section*{Acknowledgement}
This work was supported partly by the Science and Engineering Research Board (SERB) of Department of Science and Technology (DST), Government of India, through J.C Bose National Fellowship to B. Sundar Rajan.

\begin{appendices}
	\section{Proof of Corollary \ref{cor:CRD} }\label{appendix:cor:CRD}	
For any  positive integer $n \geq 2$, consider the set of points $\mathcal{X}=[n^2]$ arranged in the order shown below.
\[
\begin{bmatrix}
	1      & 2      & 3      & \cdots & n \\
	n+1    & n+2    & n+3    & \cdots & 2n \\
	2n+1   & 2n+2   & 2n+3   & \cdots & 3n \\
	\vdots & \vdots & \vdots & \ddots & \vdots \\
	(n-1)n+1 & (n-1)n+2 & (n-1)n+3 & \cdots & n^2
\end{bmatrix}
\]

\begin{figure*}[!htbp]
	\centering	
	\begin{subfigure}{0.45\textwidth}
		\centering
		\includegraphics[width=0.8\textwidth]{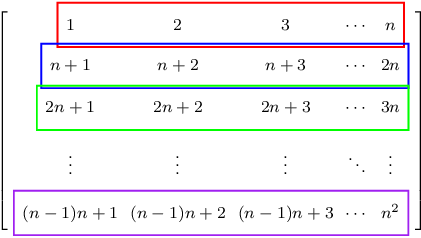}
		\subcaption{}
		\label{fig:sub1}
	\end{subfigure}
	\hfill
	\begin{subfigure}{0.54\textwidth}
		\centering
		\includegraphics[width=\textwidth]{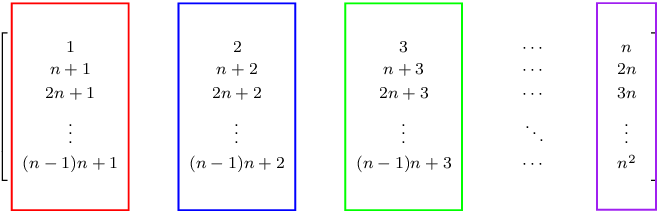}
		\subcaption{}
		\label{fig:sub2}
	\end{subfigure}
	\caption{Ways of choosing blocks $A \in \mathcal{A}$ of the MCRD mentioned in Corollary \ref{cor:CRD}}
	\label{fig:crd}
\end{figure*}

Taking the set of elements in each row as a block, as show in Fig.~\ref{fig:sub1}, gives one parallel class, whereas taking the set of elements in each column as a block, as shown in Fig.~\ref{fig:sub2}, gives another parallel class. 
From Fig.\ref{fig:crd} it is clear that  the cardinality of the intersection of two blocks drawn from two distinct parallel classes is one. i.e., $\mu_{2}=1$.
\hfill $\blacksquare$.	
	\section{Proof of Theorem \ref{thm:dpda1} }\label{appendix:dpda1}
By $(\ref{eq:dpda_constr1})$, for a given column $A_i^j$, a $\star$ appear in a row $x \in \mathcal{X}$ if and only if $x \in A_i^j $. Since $|A_i^j|=n$, the number of $\star s $ in a column $Z= n$. Thus $C1$ of DPDA definition holds.  $C2$ is obvious from the construction.

By $(\ref{eq:dpda_constr1})$, $P_{x,A_i^j} = x \cup (A_i^j \cap A^{(j \mod 2)+1}(x))$ if $ x \notin A_i^j$. For a given column $A_i^j$, all non-$\star$ entries $x \cup (A_i^j \cap A^{(j \mod 2)+1}(x))$ will be different since $x$ is distinct for each row. Thus,  $x \cup (A_i^j \cap A^{(j \mod 2)+1}(x))$ appears only once in a given column. From $(\ref{eq:dpda_constr1})$ it is clear that in a row $x \in \mathcal{X}$, $\star$ will appear only in the $2$ columns $A^j(x) : j \in [2]$, one from each of the $2$ parallel classes. The non-$\star$ entries corresponding to the remaining $(n-1) $ blocks of a parallel class $\mathcal{P}_j$, $\{A_i^j \textbackslash A^j(x): i \in [n]\}$, are the distinct $(n-1)$ sets $\{\binom{A^{(j \mod 2)+1}(x)}{2}: x \in \binom{A^{(j \mod 2)+1}(x)}{2}\}$. Since $\mu_{2}=1$, $A^j(x) \cap A^{j'}(x)=x \text{ for } j \ne j' $. Therefore, the elements of set $\binom{A^j(x)}{2}$ are different and distinct $\forall j \in [2]$. That is,  $x \cup (A_i^j \cap A^{(j \mod 2)+1}(x))$ cannot appear more than once in a row. Thus $C3.(a)$ of the DPDA definition holds.

Consider two non-$\star$ entries $P_{x,A_i^j}$, $P_{x',A_{i'}^{j'}}$ with $x \neq x'$ and $A_i^j \neq A_{i'}^{j'}$, let $ P_{x,A_i^j} = P_{x',A_{i'}^{j'}}$. Then by $(\ref{eq:dpda_constr1})$, $x \cup (A_i^j \cap A^{(j \mod 2)+1}(x))=x' \cup (A_{i'}^{j'} \cap A^{(j' \mod 2)+1}(x'))$ and $x \notin A_i^j, x' \notin A_{i'}^{j'}$. Since cardinality of non-$\star$ entry is $2$ and $x \in x \cup (A_i^j \cap A^{(j \mod 2)+1}(x))$ and $ x' \in x' \cup (A_{i'}^{j'} \cap A^{(j' \mod 2)+1}(x'))$, therefore, $x \cup (A_i^j \cap A^{(j \mod 2)+1}(x))=x' \cup (A_{i'}^{j'} \cap A^{(j' \mod 2)+1}(x'))=\{x, x'\}$. Therefore, $x'=A_i^j \cap A^{(j \mod 2)+1}(x)$ since $x \notin A_i^j$  and $x= A_{i'}^{j'} \cap A^{(j' \mod 2)+1}(x')$ since $x' \notin A_{i'}^{j'}$. This implies $x' \in A_i^j$ and $x \in A_{i'}^{j'}$. Therefore, by $(\ref{eq:dpda_constr1})$, $P_{x',A_{i}^{j}}=\star$ and $P_{x,A_{i'}^{j'}}=\star$.  Thus $C3.(b)$ of the DPDA definition holds.

We define the mapping $\phi$ from the non-$\star$ entries $\{a,b\} \in \binom{A_i^j}{2}$  to the users, denoted by the blocks $A_i^j \in \mathcal{A}$, as follows.  $\phi(\{a,b\})= A_i^j \text{ if } \{a,b\} \subseteq A_i^j $. By $(\ref{eq:dpda_constr1})$, it is clear that in the column denoted by the block $A_i^j$, there is $\star$ in all the rows denoted by the points $x \in A_i^j $ and also the row index corresponding to an entry $\{a,b\}$ is either $a$ or $b$. Therefore, the mapping  $\phi(\{a,b\})= A_i^j \text{ if } \{a,b\} \subseteq A_i^j $ ensures that $P_{x,A_i^j}=\star$ where $x \in \{a,b\}$. Thus $C4$ of the DPDA definition holds.
	\hfill $\blacksquare$.
	\section{Proof of Corollary \ref{cor:conc1_gen} }\label{appendix:dpda1_gen}
	Consider a CRD $(\mathcal{X}, \mathcal{A})$ with |$\mathcal{X}$| = $v$, |$\mathcal{A}$| = $b$ , block size $k$, $r=\frac{bk}{v}$ parallel classes and $\mu_{2}=1$. We denote block $A \in \mathcal{A}$ as $A_i^j : i \in [n], j \in [r]$, where $A_i^j$ denotes the $i^{th}$ block in $j^{th}$ parallel class $\mathcal{P}_j$. We use the notation $A^j(x)$ to denote block $A$ in the $j^{th}$ parallel class to which the point $x$ belongs. We construct an array $\mathbf{P}$  whose rows are indexed by all the points $x \in \mathcal{X}$ and columns are indexed by all the blocks $A_i^j \in \mathcal{A}$, defined as follows.   	
	\begin{equation}\label{eq:dpda_constr1_gen}
		P_{x,A_i^j}= \begin{cases}   \hspace{1cm}\star, & \text {if } x \in A_i^j    \\  x \cup (A_i^j \cap A^{(j \mod r)+1}(x)), & \text {otherwise }\end{cases}.
	\end{equation}
	The proof that (\ref{eq:dpda_constr1_gen}) results in the DPDA specified in Corollary \ref{cor:conc1_gen} is the same as the proof of Theorem \ref{thm:dpda1} except $|\mathcal{X}| = v$, $|\mathcal{A}| = b$, $|\mathcal{A_i^j}| = k$ and replacing $A^{(j \mod 2)+1}(x)$ in the proof by $A^{(j \mod r)+1}(x)$. 
	\section{Proof of Theorem \ref{thm:dpda4} }\label{appendix:dpda4}
	By $(\ref{eq:dpda_constr4})$, for a given column $x \in \mathcal{X}$, a $\star$ appear in a row $A \in \mathcal{A}$ if $x \notin A$. By Theorem \ref{thm:tgdd}, the number of blocks in a $2-(n,2,2,1)$ GDD in which a given point $x$ belongs is $2(n-1)$. Therefore, the number of blocks to which $x$ does not belong is $2n(n-1)-2(n-1)=2(n-1)^2$. That is, $Z= 2(n-1)^2$. Thus $C1$ of the DPDA definition holds.  $C2$ is obvious from the construction.
	
	By $(\ref{eq:dpda_constr4})$, $P_{A,x} = A\backslash \{x\}$ if $ x \in A$. Consider a given column indexed by $x$. Let $\mathcal{A}(x)=\{A_1(x), A_2(x),...,A_{2(n-1)}(x)\}$ denotes the set of blocks in $\mathcal{A}$ in which $x$ belongs. The blocks $A_i(x)$ are of the form $\{x,x_i\}$, where $x_i \ne x$. By the definition of a $t$-GDD, for $x_i \in A_i(x)$ and $x_{i'} \in A_{i'}(x)$, $x_i \ne x_{i'}$. Therefore, $A\backslash \{x\}$  only appears once in a given column. From $(\ref{eq:dpda_constr4})$, the non-$\star$ entries in a given row $A \in \mathcal{A}$ are the points in $A$, which are distinct. Therefore,  $A\backslash \{x\}$ cannot appear more than once in a row. Thus $C3.(a)$ of the DPDA definition holds.
	
	Consider two non-$\star$ entries $P_{A,x}$, $P_{A',x'}$ with $A \neq A'$ and $x \neq x'$. Let $ P_{A,x} = P_{A',x'}$. Then by $(\ref{eq:dpda_constr4})$, $A\backslash \{x\}=A'\backslash \{x'\}$. Since $|A|=|A'|=2$, let $A\backslash \{x\}=A'\backslash \{x'\}=y$, for some $y \in \mathcal{X}$.  Since $x,y \in A$ and $|A|=2$, $x' \notin A$. Similarly $x \notin A'$. Therefore by $(\ref{eq:dpda_constr4})$, $P_{x',A} =\star$ and $P_{x,A'} =\star$. Thus $C3.(b)$ of the DPDA definition holds.
	
	We define the mapping $\phi$ from the non-$\star$ entries $A\backslash \{x\}=\{y\}$  to the users, denoted by $\{x\} \in \mathcal{X}$, as $\phi(\{y\})= \{x\} : x,y \in G_i, \text{for some } G_i \in \mathcal{G} $. Since $x$ and $y$ belong to the same group $G_i$, by definition of $t$-GDD, the blocks in $\mathcal{A}(x)$ do not contain the point $\{y\}$. Therefore by $(\ref{eq:dpda_constr4})$, $P_{\mathcal{A}(x),y} =\star$. Since $|G_i|=2 \text{ } \forall i \in [n]$, for every $y \in \mathcal{X}$, there is a unique $x \in \mathcal{X}$ such that $P_{\mathcal{A}(x),y} =\star$.  Thus $C4$ of the DPDA definition holds. \hfill $\blacksquare$.
	\section{Proof of Theorem \ref{thm:dpda5} }\label{appendix:dpda5}
	By $(\ref{eq:dpda_constr5})$, for a given column $z \in [n]$, a $\star$ appear in a row $(x,y) \in \mathcal{S}_n'$ if $z \notin \{x,y\}$. By definition of $\mathcal{S}_n'$, each integer $z$ appears as the first entry in exactly $n-2$ ordered pairs in $\mathcal{S}_n'$ and as the second entry in exactly another $n-2$ ordered pairs. Therefore, the number of ordered pairs in $\mathcal{S}_n'$ in which $z$ does not belong is $n(n-2)-2(n-2)=(n-2)^2$. That is, $Z= (n-2)^2$. Thus $C1$ of the DPDA definition holds.  $C2$ is obvious from the construction.
	
	By $(\ref{eq:dpda_constr5})$, $P_{(x,y),z}= (\{x,y\} \backslash z)_\alpha$ if $ z \in \{x,y\}$. Therefore, in a given column indexed by $z$, the non-$\star$ entries appear only in those rows that are indexed by either $(z,y)$ or $(x,z)$, where $x,y \ne z$, and the corresponding entries are $y_1$ or $x_2$, respectively. Therefore, when $x$ and $y$ are the same, the corresponding $\alpha$ value is different and for the same $\alpha$ value $x$ or $y$ will not appear twice since no ordered pair in $\mathcal{S}'_n$ appears twice. Therefore, a specific non-$\star$ entry  $(\{x,y\} \backslash z)_\alpha$ in a given column appears only once. In a given row indexed by an ordered pair $(x,y)$, the non-$\star$ entries will be either $x_\alpha$ or $y_\alpha$, for some $\alpha \in [2]$. Since $x \ne y$, a non-$\star$ entry cannot appear more than once in a row. Thus $C3.(a)$ of the DPDA definition holds.
	
	Consider two non-$\star$ entries $P_{(x,y),z}$ and $P_{(x',y'),z'}$ with $(x,y) \neq (x',y')$ and $z \neq z'$. Let $ P_{(x,y),z} = P_{(x',y'),z'}$. If $\alpha=1$ in these entries, then, by $(\ref{eq:dpda_constr5})$, $z=x, z'=x'$ and $y=y' \notin \{z,z'\}$. Since $(x,y) \neq (x',y')$ and $y=y'$, $x \ne x'$. Thus, $x \ne z'$. If $\alpha=2$ in the considered entries, then, by $(\ref{eq:dpda_constr5})$, $z=y, z'=y'$ and $x=x' \notin \{z,z'\}$. Since $(x,y) \neq (x',y')$ and $x=x'$, $y \ne y'$. Thus, $y \ne z'$. Thus in both cases of $\alpha$ values, $z' \notin \{x,y\}$. Therefore, by $(\ref{eq:dpda_constr5})$,  $P_{(x,y),z'}=\star$. Similarly $P_{(x',y'),z}=\star$. Thus $C3.(b)$ of the DPDA definition holds.
	
	We define the mapping $\phi$ from the non-$\star$ entries $p_\alpha$ to the users $[K]$ denoted by $p$, where $p \in (1,2,...,n)$ as follows.
	\begin{equation*}
		\phi\left(p_\alpha\right)= \begin{cases}    ((p-2) \text{ mod } n) + 1 & \text { if } \alpha = 1   \\ (p \text{ mod } n ) + 1 & \text { if } \alpha = 2. \end{cases}
	\end{equation*}
	That is, when $\alpha=1$ and $\alpha=2$, $\phi\left(p_\alpha\right)$ is the nearest left and nearest right entries of $p$ in a cyclic fashion in the ordered $n$-tuple $(1,2,...,n)$, respectively, $\forall p \in [n]$. First, we consider the case $\alpha=1$. Let $P_{(x,y),z}=p_1$. Then, by $(\ref{eq:dpda_constr5})$, $z=x$, $y=p$ and $x \ne p$. In the construction of the DPDA, the rows are indexed by all the ordered pairs in $\mathcal{S}'_n$, which does not contain any ordered pairs  \{$(1,2)$, $(2,3)$, $....$, $(n-1,n)$, $(n,1)$\}. Therefore, $x \ne ((p-2) \text{ mod } n) + 1$ for any $p \in (1,2,..,n)$ when $P_{(x,y),z}=p_1$. Therefore, $P_{(x,y),((p-2) \text{ mod } n) + 1}=\star$. Now consider the case $\alpha=2$. Let $P_{(x,y),z}=p_2$. Then, by $(\ref{eq:dpda_constr5})$, $z=y$, $x=p$ and $y \ne p$. And since $\mathcal{S}'_n$ does not contain any ordered pairs  \{$(1,2)$, $(2,3)$, $....$, $(n-1,n)$, $(n,1)$\}, $y \ne (p \text{ mod } n) + 1$ for any $p \in (1,2,..,n)$ when $P_{(x,y),z}=p_2$. Therefore, $P_{(x,y),(p \text{ mod } n) + 1}=\star$. Thus $C4$ of the DPDA definition holds. \hfill $\blacksquare$.
			\section{Proof of Theorem \ref{thm:dpda2} }\label{appendix:dpda2}
	By $(\ref{eq:dpda_constr2})$, for a given column $x$, a $\star$ appear in a row $A_i^j \in \mathcal{A}$ if and only if $x \in A_i^j $. Since there are two parallel classes, and in each parallel class there is exactly one block that contains the point $x$, and the number of $\star$s in a column is $Z= 2$. Thus $C1$ of the DPDA definition holds.  $C2$ is obvious from the construction.
	
	By $(\ref{eq:dpda_constr2})$, $P_{A_i^j,x} = (A_i^j \cap A^{(j \mod 2)+1}(x))_{\alpha}$ if $ x \notin A_i^j$. Consider a given column indexed by point $x$. For this column, from $(\ref{eq:dpda_constr2})$ it is clear that non-$\star$ entries will be in all the columns except those denoted by $A^j(x), j \in [2]$. Since $A_i^j$ does not contain $x$, $A_i^j \cap A^{(j \mod 2)+1}(x) \subseteq  A^{(j \mod 2)+1}(x)$ and $|(A_i^j \cap A^{(j \mod 2)+1}(x))|=1$, the non-$\star$ entries in the $(n-1)$ rows denoted by blocks of a specific parallel class $\mathcal{P}_j$ are the $(n-1)$ different points in the block $A^{(j \mod 2)+1}(x)$. Therefore  $A_i^j \cap A^{(j \mod 2)+1}(x)$ will be different in a column denoted by $x$. Thus   $(A_i^j \cap A^{(j \mod 2)+1}(x))_{\alpha}$ appear only once in a given column.  Since $\alpha$ denotes the $\alpha^{th}$ occurrence of $A_i^j \cap A^{(j \mod 2)+1}(x)$ from left to right in the row indexed by $A_i^j$, $(A_i^j \cap A^{(j \mod 2)+1}(x))_{\alpha}$ cannot appear more than once in a row. Thus $C3.(a)$ of the DPDA definition holds.
	
	Consider two non-$\star$ entries $P_{A_i^j,x}$, $P_{A_{i'}^{j'},x'}$ with $x \neq x'$ and $A_i^j \neq A_{i'}^{j'}$, let $ P_{A_i^j,x} = P_{A_{i'}^{j'},x'}$. Then by $(\ref{eq:dpda_constr2})$, $(A_i^j \cap A^{(j \mod 2)+1}(x))=(A_{i'}^{j'} \cap A^{(j' \mod 2)+1}(x'))$ and $x \notin A_i^j, x' \notin A_{i'}^{j'}$. Let $(A_i^j \cap A^{(j \mod 2)+1}(x))=(A_{i'}^{j'} \cap A^{(j' \mod 2)+1}(x'))=y \in \mathcal{X}$ since $|(A_i^j \cap A^{(j \mod 2)+1}(x))|=1$. Then clearly $y \in A_i^j$ and $y \in A_{i'}^{j'}$. Also $y \in A^{(j \mod 2)+1}(x)$ and $y \in A^{(j' \mod 2)+1}(x')$. Since a point $y$ can occur exactly in one block in a parallel class, $A_i^j = A^{(j' \mod 2)+1}(x')$ and $A_{i'}^{j'}=A^{(j \mod 2)+1}(x)$.  That implies $x' \in A_i^j$ and $x \in A_{i'}^{j'}$. Therefore by $(\ref{eq:dpda_constr2})$, $P_{A_i^j,x'} =\star$ and $P_{A_{i'}^{j'},x} =\star$. Thus $C3.(b)$ of the DPDA definition holds.
	
	We define the mapping $\phi$ from the non-$\star$ entries $\{ a_\alpha \} : a \in \mathcal{X}$  to the users $[K]$ denoted by the points in $\mathcal{X}$ as an identity function.  That is, $\phi(\{a_\alpha \})= a $. By $(\ref{eq:dpda_constr2})$, it is clear that the $a$ in the non-$\star$  entry $a_\alpha$  in a row $A_i^j$ is a point in $A_i^j$ and the column corresponding to the user denoted by $a$ is $\star$ since $a \in  A_i^j$. Thus $C4$ of the DPDA definition holds. \hfill $\blacksquare$. 
		\section{Proof of Theorem \ref{thm:dpda3} }\label{appendix:dpda3}
		By $(\ref{eq:dpda_constr3})$, for a given column $\mathcal{C}$, a $\star$ appear in a row $i$ if and only if $i \in \mathcal{C}$. Since $|\mathcal{C}|=k$, the number of $\star$s in a column is $Z= k$. Thus $C1$ of the DPDA definition holds.  $C2$ is obvious from the construction.
		
		By $(\ref{eq:dpda_constr3})$, $P_{i,\mathcal{C}} = (i \cup \mathcal{D}(i,\mathcal{C}))_{\alpha}$ if $ i \notin \mathcal{C}$. Consider a given column indexed by the $\mathcal{C}$. For this column, from $(\ref{eq:dpda_constr3})$ it is clear that non-$\star$ entries will be in all the rows indexed by $i$ such that $ i \notin \mathcal{C}$. Since $ i \notin \mathcal{C}$ and the non-$\star$ entry $P_{i,\mathcal{C}} = (i \cup \mathcal{D}(i,\mathcal{C}))_{\alpha}$, $(i \cup \mathcal{D}(i,\mathcal{C}))$ appear only once in a given column $\mathcal{C}$. Thus $(i \cup \mathcal{D}(i,\mathcal{C}))_{\alpha}$  appear only once in a given column. Since $\alpha$ denotes the $\alpha^{th}$ occurrence of $(i \cup \mathcal{D}(i,\mathcal{C}))$ from left to right in the row indexed by $i$, $(i \cup \mathcal{D}(i,\mathcal{C}))_{\alpha}$ cannot appear more than once in a row. Thus $C3.(a)$ of the DPDA definition holds.
		
		Consider two non-$\star$ entries $P_{i,\mathcal{C}}$, $P_{i',\mathcal{C}'}$ with $i \neq i'$ and $\mathcal{C} \neq \mathcal{C}'$. Let $ P_{i,\mathcal{C}} = P_{i',\mathcal{C}'}$. Then by $(\ref{eq:dpda_constr3})$, $(i \cup \mathcal{D}(i,\mathcal{C}))=(i' \cup \mathcal{D}'(i',\mathcal{C}'))$, where $\mathcal{D}(i,\mathcal{C}) \subset \mathcal{C}$ and $\mathcal{D}'(i',\mathcal{C}') \subset \mathcal{C}'$. This implies $i \in \mathcal{D}'(i',\mathcal{C}')$ and $i' \in \mathcal{D}(i,\mathcal{C})$. Therefore, $i \in \mathcal{C}'$ and $i' \in \mathcal{C}$. Therefore by $(\ref{eq:dpda_constr3})$, $P_{i,\mathcal{C}'} =\star$ and $P_{i',\mathcal{C}} =\star$. Thus $C3.(b)$ of the DPDA definition holds.
		
		We define the mapping $\phi$ from the non-$\star$ entries $(i \cup \mathcal{D}(i,\mathcal{C}))_{\alpha}$  to the users $[K]$ denoted by $\mathcal{C}$ as $\phi((i \cup \mathcal{D}(i,\mathcal{C}))_{\alpha})= \mathcal{C} : (i \cup \mathcal{D}(i,\mathcal{C}))=\mathcal{C} $. Since $(i \cup \mathcal{D}(i,\mathcal{C}))=\mathcal{C}$, $i \in \mathcal{C}$ and therefore by $(\ref{eq:dpda_constr3})$, $P_{i,\mathcal{C}} =\star$. Thus $C4$ of the DPDA definition holds.
		\hfill $\blacksquare$. 
\end{appendices}


\begin{thebibliography}{1}	
		\bibitem{Eri}
		Ericsson mobility report-June 2024. Available on:\\ https://www.ericsson.com/en/reports-and-papers/mobility-report/reports/june-2024
		
			\bibitem{MaN}
		M. A. Maddah-Ali and U. Niesen, "Fundamental Limits of Caching," in IEEE Transactions on Information Theory, vol. 60, no. 5, pp. 2856-2867, May 2014.
		
		\bibitem{YMA}
		Q. Yu, M. A. Maddah-Ali and A. S. Avestimehr, "The Exact Rate-Memory Tradeoff for Caching With Uncoded Prefetching," in IEEE Transactions on Information Theory, vol. 64, no. 2, pp. 1281-1296, Feb. 2018.
		
		\bibitem{ERW}
		B. M. Elhalawany, R. Ruby and K. Wu, "D2D Communication for Enabling Internet-of-Things: Outage Probability Analysis," in IEEE Transactions on Vehicular Technology, vol. 68, no. 3, pp. 2332-2345, March 2019.
		
		\bibitem{MACF}
		L. Militano, G. Araniti, M. Condoluci, I. Farris, and A. Iera, , "Device-to-device communications for 5G internet of things," EAI Endorsed Transactions on Internet of Things, vol. 1, no. 1, pp.1-15, Oct. 2015.
		
		\bibitem{LLLW}
		Y. Li, Y. Liang, Q. Liu and H. Wang, "Resources Allocation in Multicell D2D Communications for Internet of Things," in IEEE Internet of Things Journal, vol. 5, no. 5, pp. 4100-4108, Oct. 2018.
		
		\bibitem{TUY}
		M. N. Tehrani, M. Uysal and H. Yanikomeroglu, "Device-to-device communication in 5G cellular networks: challenges, solutions, and future directions," in IEEE Communications Magazine, vol. 52, no. 5, pp. 86-92, May 2014.
		
		\bibitem{ZTS}
		I. Zyrianoff, A. Trotta, L. Sciullo, F. Montori and M. Di Felice, "IoT Edge Caching: Taxonomy, Use Cases and Perspectives," in IEEE Internet of Things Magazine, vol. 5, no. 3, pp. 12-18, September 2022.
		
		\bibitem{Ji}
		M. Ji, G. Caire and A. F. Molisch, "Fundamental Limits of Caching in Wireless D2D Networks," in IEEE Transactions on Information Theory, vol. 62, no. 2, pp. 849-869, Feb. 2016.
		
		\bibitem{CKRG}
		Ç. Yapar, K. Wan, R. F. Schaefer and G. Caire, "On the Optimality of D2D Coded Caching With Uncoded Cache Placement and One-Shot Delivery," in IEEE Transactions on Communications, vol. 67, no. 12, pp. 8179-8192, Dec. 2019.
		
		\bibitem{YCT}
		Q. Yan, M. Cheng, X. Tang and Q. Chen, "On the Placement Delivery Array Design for Centralized Coded Caching Scheme," in IEEE Transactions on Information Theory, vol. 63, no. 9, pp. 5821-5833, Sept. 2017.
		
		\bibitem{JMQX}
		J. Wang, M. Cheng, Q. Yan and X. Tang, "Placement Delivery Array Design for Coded Caching Scheme in D2D Networks," in IEEE Transactions on Communications, vol. 67, no. 5, pp. 3388-3395, May 2019.
		
		\bibitem{JMQX_Arxiv}
		J. Wang, M. Cheng, Q. Yan and X. Tang, "Placement Delivery Array Design for Coded Caching Scheme in D2D Networks," Available on arXiv: 1712.06212 [cs.IT], Dec 2017.
		
		\bibitem{WCJ_RS}
		N. Woolsey, R. -R. Chen and M. Ji, "Device-to-Device Caching Networks with Subquadratic Subpacketizations," GLOBECOM 2017 - 2017 IEEE Global Communications Conference, Singapore, 2017, pp. 1-6.
		
		\bibitem{WCJ_HY}
		N. Woolsey, R. -R. Chen and M. Ji, "Coded Caching in Wireless Device-to-Device Networks Using a Hypercube Approach," 2018 IEEE International Conference on Communications Workshops (ICC Workshops), Kansas City, MO, USA, 2018, pp. 1-6.
		
		\bibitem{JY}
		J. Li and Y. Chang, "New Constructions of D2D Placement Delivery Arrays," in IEEE Communications Letters, vol. 27, no. 1, pp. 85-89, Jan. 2023.
		
		\bibitem{MADCC_arxiv}
		Rashid Ummer N.T., and B. Sundar Rajan. "D2D Coded Caching Schemes for Multiaccess Networks with Combinatorial Access Topology", 2025 IEEE International Symposium on Information Theory (ISIT), Ann Arbor, MI, USA, 2025, pp. 1-6.
		
		\bibitem{Col}
		C. J. Colbourn and J. H. Dinitz, Handbook of combinatorial designs. CRC press, 2006.
		
		\bibitem{DPB}
		D. Katyal, P. N. Muralidhar and B. S. Rajan, "Multi-Access Coded Caching Schemes From Cross Resolvable Designs," in IEEE Transactions on Communications, vol. 69, no. 5, pp. 2997-3010, May 2021.
		
		\bibitem{HM}
		Mohácsy, Hedvig. "The asymptotic existence of group divisible designs of large order with index one", Journal of Combinatorial Theory, Series A 118, no.7, pp.1915-1924, 2011.
		
		\bibitem{Stin}
		Stinson, Douglas Robert. Combinatorial designs: constructions and analysis. Vol. 480. New York: Springer, 2004.
		
		\bibitem{HRW}
		Hanani, H., Ray-Chaudhuri, D.K. and Wilson, R.M., On resolvable designs. Discrete Mathematics, 3(4), pp.343-357, 1972.
		
		\bibitem{TaR}
		L. Tang and A. Ramamoorthy, "Coded Caching Schemes With Reduced Subpacketization From Linear Block Codes," in IEEE Transactions on Information Theory, vol. 64, no. 4, pp. 3099-3120, April 2018.
		
		\bibitem{SHMWT}
		L. Sijia, Y. Han, L. Ma, L. Wang, and Z. Tian. "A generalization of group divisible t‐designs." Journal of Combinatorial Designs 31, no. 11, pp.575-603, 2023.
		
		\bibitem{cyc_des}
		John, Joshua A., and J. A. John. Cyclic designs. Springer US, 1987.
		
		
	\end{thebibliography}
\end{document}